\documentclass[final,3p,times]{elsarticle}

\usepackage{amssymb}
\usepackage{lineno}
\usepackage[T1]{fontenc}   
\usepackage{textcomp}
\usepackage{subfig}
\usepackage{graphicx}
\usepackage{amsmath}
\usepackage{bm}
\usepackage[version=4]{mhchem}
\usepackage{siunitx}
\usepackage{longtable,tabularx}
\usepackage{color}
\usepackage{color}
\usepackage[dvipsnames]{xcolor}
\usepackage{tcolorbox}
\usepackage{hyperref}
\usepackage{multirow}
\usepackage{enumitem}
\usepackage{wrapfig}

\colorlet{colorRev1}{blue!80!black} 
\colorlet{colorRev2}{red!80!black} 

\begin{document}
	
	\begin{frontmatter}

	\title{Integrated modelling of microvibrations induced by Solar Array Drive Mechanism for worst-case end-to-end analysis and robust disturbance estimation}
	
	\author[1]{Francesco Sanfedino\corref{cor1}}
	\ead{francesco.sanfedino(at)isae.fr}
	\author[1]{Daniel Alazard}
	\author[2]{Valentin Preda}
	\author[2]{Davide Oddenino}

    \address[1]{Institut Sup\'erieur
    de l'A\'eronautique et de l'Espace (ISAE-SUPAERO), Toulouse, France}
	\address[2]{European Space Agency (GNC, AOCS \& Pointing Division), ESTEC, Noordwijk, The Netherlands}
		
	\cortext[cor1]{Corresponding author}	
	
	\begin{abstract}	    
Modern and future observation Space missions face increasingly more demanding pointing performance requirements. This is accompanied with the development of larger lightweight flexible structures. This paper outlines a methodology for modeling a generic multi-body flexible spacecraft in an end-to-end fashion within the Two-Inputs Two-Outputs Port framework: from disturbance all the way to pointing performance by tacking into account all the uncertainties of each sub-system. A particular focus is dedicated to a novel generic model for Solar Drive Array Mechanisms and its harmonic disturbances produced by the micro-stepping driver and the gearbox imperfections.  The proposed model is validated with the on-board telemetries of a European spacecraft. Moreover this paper shows how the proposed model-based approach can be easily used for control design and closed-loop robust performance analysis. Worst-case scenarios of mechanical parameters combination are investigated to provide robust performance certificate of the reference study case. A Linear Parameter-Varying observer, scheduled by the solar array rotor angle, is finally proposed in order to estimate the disturbance torques induced by the gearbox imperfections just using the measurements coming from a classical attitude control system: a star tracker and a gyrometer. 

	\end{abstract}


	\end{frontmatter}

    \section{Introduction}

\subsection{Background and motivation}

In the last decade the European Space Agency (ESA) and NASA made an effort to join their experience to tackle in a more systematic way the problem of fine pointing for the next generation high accuracy observation and Science missions \cite{denney2018}.

This problem offers a clear benchmark of multi-disciplinary nature: structure, control and system engineering are involved in order to limit as much as possible the propagation and amplification of the internal disturbance signals through the flexible structure of the spacecraft. 
This is why it is crucial to develop rigorous methodology to model and predict worst-case scenario to avoid mission requirement degradation as done in the recent works \cite{Preda2017a,sanfe2019} in early preliminary design phase. 
However the biggest difficulty in preventing in-flight model mismatches is to obtain high fidelity models before flying by comparing analytical and numerical Finite Element models (FEM) with identification experiments on ground and in presence of gravity. The Solar Dynamics Observatory (SDO) represents an example of this difficult task \cite{Liu2007}. The on-ground test of disturbance to payload line-of-sight assessment was not performed because of the intrinsic difficulty of reproducing the micro-gravity conditions. Afterwords a deep on-orbit jitter analysis revealed the performance limitations to impose to the reaction wheel assembly and the  high gain antenna (HGA) stepper motors in order to mitigate the induced micro-vibrations. 

What is extremely difficult to capture in on-ground tests is the effective dynamic coupling effect among the different spacecraft bodies and the actual damping of the various structural modes \cite{denney2018}. What is more is that generally the dynamical characteristics tend to evolve during the mission life-cycle by shifting natural frequencies and by varying the modal amplitudes. 

The widespread practice for jitter predictions is to use FEM models with a high number of states. Nonetheless this task results hard if a FEM model has to be provided for each predicted reaction wheel speed or a different angular configuration of a solar panel driven by a Solar Array Drive Mechanism (SADM). 
As stated in \cite{denney2018} for BepiColombo mission a Monte Carlo campaign was discarded for jitter analysis since a different FEM model should have been considered for any combination of HGA and Solar Array (SA) angular configurations. The adopted approach was then to extract a confidence interval by extracting data from only seven scenarios. This method was not able to provide any worst-case scenario for the HGA and SA configurations together with the worst-case mechanism/wheel speeds.

Linear Fractional Transformation (LFT) and structured singular values framework \cite{KEMINZHOU1996} represent nowadays robust and powerful tools to tackle this difficult task in an alternative way by considering in one model all the system uncertainties and parameter variations. Furthermore they allow reducing the stability/performance analysis time and can detect worst-case scenario without relaying on classical non-global time-consuming simulation-based approaches like Monte Carlo campaigns.

In the last ten years the \textit{Institut Superieur de l'Aéronautique et de l'Espace} (ISAE-SUPAERO) works on the development of minimal LFT models of flexible spacecrafts. The developed framework is a substructure synthesis approach called Two-Input Two-Output Ports (TITOP) \cite{Alazard2015,Perez2015,Perez2016,Chebbi2017,sanfedino2018finite} that consists in connecting several flexible sub-structures through dynamical ports by keeping the uncertain nature of the plant and condensing all the possible scenarios (variation of the mechanical properties and geometrical configurations, variation of reaction wheel speed, etc.) in a unique LFT model. The peculiarity of the TITOP theory with respect to similar substructure approaches in literature as \cite{AN2020106934,LIU2021107264} consists in providing dynamical models directly adapted for robust performance certification and robust control synthesis.  

All the models derived in TITOP approach have been recently implemented in the last release of the Satellite Dynamics Toolbox (SDT) \cite{alazard2020,al2021stdlib,alaSanf2021}, which allows the user to easily build the model of a flexible spacecraft with several appendages by assembling elemental \textsc{Simulink} customized blocks.
This model is then ready for robust control synthesis and robust stability and performance assessment \cite{sanfe2019,sanfedino2020robust,preda2020robust} by using the available \textsc{Matlab} routines of the \textit{Robust Control Toolbox} \cite{balas2016}.  

The present work aims at presenting a novel model of SADM system together with its related disturbing signals in the TITOP framework. 
Charbonnel \textit{et al.} \cite{Valentin-Charbonnel1999} presented an $\mathcal{H}_\infty$ attitude control design for an observation satellite by tacking into account both reaction wheel and SADM disturbances. Cumer \textit{et al.} \cite{Chretien2001} provided a minimal LFT form of a flexible spacecraft with a SADM.
Beugnon \textit{et al.} \cite{beugnon2003mixed} presented worst-case stability assessment of the MetOp spacecraft.
In all these works the SADM is modeled as an equivalent mass-spring mechanical system whose parameters values do not explicitly depend on the motor characteristics.

Cao \textit{et al.} \cite{CAO2019131} proposed a sliding mode control and input shaping technique for compensation of SADM speed fluctuation.
A  phase compensation active control strategy is proposed by Guo \textit{et al.} \cite{GUO2020106148} in order to reduce the attitude disturbances resulting from SADM. Zhang \textit{et al.} \cite{ZHANG2021106398} proposed an original active suppression strategy for SADM drive fluctuations. 
In all these works the SADM model is finely detailed but any formal certificate of performance robustness is provided.

To the best of the authors' knowledge, no prior work in literature presented a detailed SADM dynamic model that is suitable for end-to-end robust control design and fine worst-case analysis. In the proposed framework, the impact of the dynamics and perturbations introduced by the SADM to various performance and stability metrics can be indeed accurately analyzed in 6 degrees of freedom case.

The first contribution of this paper is to propose a novel Linear Parameter-Varying (LPV) model of SADM inspired by \cite{Chen2016,Chen2017} in TITOP framework. 

The second contribution consists in presenting a methodology to model the disturbance coming from the SADM micro-stepper driver as well as from the noise generated within the teeth imperfections of the reduction gearbox for accurate robust pointing performance assessment. This work was motivated in fact by some anomalies observed in the telemetries of a European observation spacecraft which have been successfully correlated to the SADM gearbox gear teeth imperfections thanks to the detailed description of the mechanism provided by this research.
The SADM model with its perturbation signals is thus validated with the recorded telemetry flight data of the cited European spacecraft.

As third contribution an original LPV observer scheduled on the solar array angular configuration is proposed. This observer based on the SADM TITOP model is able to estimate the SADM disturbances directly on board by using the classical Attitude Control System (ACS) measurements coming from a star tracker and a gyrometer of an Inertial Measurement Unit (IMU).

After presenting the TITOP approach with some basics blocks, Section \ref{sec:problem_formulation_and_system_modeling} is dedicated to the TITOP model of a SADM (stepper motor and reduction gearbox).
SADM harmonic disturbances are modeled in Section \ref{sec:SADM_disturbance} from three different sources: micro-stepping driver, detent motor torque and gearbox gear teeth imperfections.
All the developed models are then validated with the on-board telemetries of a European spacecraft both in time and frequency domain in Section \ref{sec:model_validation}. 
The spacecraft TITOP dynamical model is finally used for worst-case robust  analysis in Section \ref{sec:wc_pointing_analysis} to asses the pointing performance in presence of a set of parametric uncertainties. Section \ref{sec:observer} outlines the LPV observer for on-board estimation of the induced disturbances.
Section \ref{sec:conclusion} finally summarizes all the achieved results.

    \section{Problem formulation and system modeling}
\label{sec:problem_formulation_and_system_modeling}

\subsection{Summary on the TITOP model approach}
\label{sec:summary_on_the_titop_model_approach}
 
\begin{figure*}[th]
\centering
\begin{minipage}{0.65\columnwidth}
\centering
	\includegraphics[width=\columnwidth]{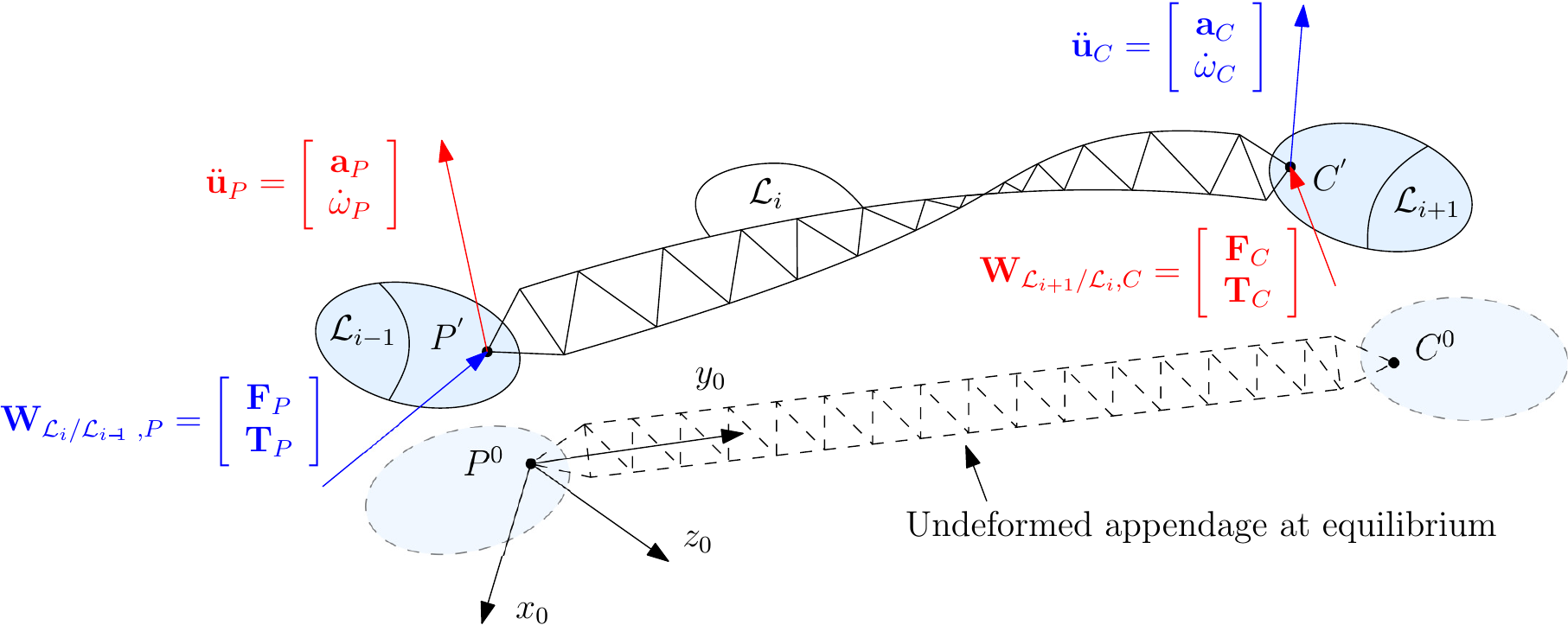}
\end{minipage}\quad
\begin{minipage}{0.27\columnwidth}
\centering
	\includegraphics[width=\columnwidth]{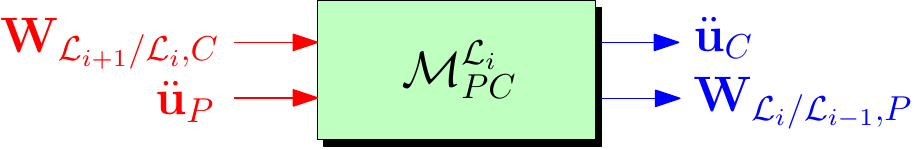}
\end{minipage}
\begin{minipage}{0.7\columnwidth}
\caption{$i$-th flexible appendage of a complex sub-structured body.}
\label{fig:TITOP}
\end{minipage}
\begin{minipage}{0.27\columnwidth}
\caption{$\mathbf{\mathcal{M}}_{PC}^{\mathcal{L}_i}$ block-diagram.}
\label{fig:block_titop}
\end{minipage}
\end{figure*}

The TITOP model $\mathbf{\mathcal{M}}_{PC}^{\mathcal{L}_i}$ of a flexible appendage $\mathcal{L}_i$ connected to a parent substructure $\mathcal{L}_{i-1}$ at point $P$ and to a child substructure $\mathcal{L}_{i+1}$ at point $C$ (see Fig. \ref{fig:TITOP}) is a linear state-space model with twelve inputs (six for each of the two input ports):
\begin{enumerate}[noitemsep,topsep=0pt]
\item The 6 components in $\mathcal{R}_0$ of the wrench $\mathbf{W}_{\mathcal{L}_{i+1}/\mathcal{L}_{i},C}$ composed of the three-components force vector $\mathbf{F}_C$ and the three-components torque vector $\mathbf{T}_C$ applied by $\mathcal{L}_{i+1}$ to $\mathcal{L}_{i}$ at the free node $C$;
\item The 6 components in $\mathcal{R}_0$ of the acceleration vector $\ddot{\mathbf{u}}_P$ composed of the three-components linear acceleration vector $\mathbf{a}_P$ and the three-components angular acceleration vector $\dot{\bm{\omega}}_P$ at the clamped node $P$;
\end{enumerate}
and twelve outputs (6 for each of the two output ports):
\begin{enumerate}[noitemsep,topsep=0pt]
\item The 6 components in $\mathcal{R}_0$ of the acceleration vector $\ddot{\mathbf{u}}_C$ at the free node $C$;
\item The 6 components in $\mathcal{R}_0$ of the wrench $\mathbf{W}_{\mathcal{L}_{i}/\mathcal{L}_{i-1},P}$ applied by $\mathcal{L}_{i}$ to the parent structure $\mathcal{L}_{i-1}$ at the clamped node $P$.
\end{enumerate}
The block diagram of the TITOP model $\mathbf{\mathcal{M}}_{PC}^{\mathcal{L}_i}$ is illustrated in Fig. \ref{fig:block_titop}.

\subsection{N-port model of a rigid central body}
\label{sec:model_central_body}

\begin{wrapfigure}[12]{r}{0.4\columnwidth}
	\begin{center}
		\vspace{-20pt}
		\includegraphics[width=.35\columnwidth]{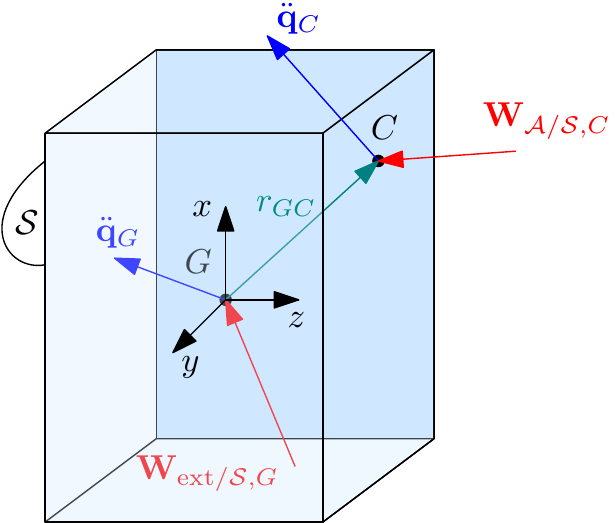}
	\end{center}
	\vspace{-10pt}
	\caption{Spacecraft body diagram.}
	\label{fig:satellite_dynamics}
\end{wrapfigure}

A satellite can be modeled as a central rigid body $\mathcal{S}$ to which several flexible appendages $\mathcal{A}_{\bullet}$ (solar panel, antenna, robotic arm, etc.) are attached. Let us consider the spacecraft $\mathcal{S}$ in Fig. \ref{fig:satellite_dynamics} with center of mass $G$. An appendage $\mathcal{A}$ is attached to the main hub at the point $C$. The objective is to write the TITOP model of $\mathcal{S}$ where the inputs are the wrenches $\mathbf{W}_{\mathrm{ext}/\mathcal{S},G}\in \mathbb{R}^{6\times 1}$ of the external forces/torques acting on the spacecraft at its center of mass $G$ (i.e. solar pressure, gravity gradient, aerodynamic drag for Low Earth Orbits, etc.) and $\mathbf{W}_{\mathcal{A}/\mathcal{S},C}\in \mathbb{R}^{6\times 1}$ of the forces/torques imposed by the appendage to the hub at point $C$. The outputs of the model are the accelerations $\ddot{\mathbf{q}}_G$ and $\ddot{\mathbf{q}}_C$ of the points $G$ and $C$ respectively.

The dynamics of the spacecraft $\mathcal{S}$ at its center of mass $G$ is given by the classic Newton-Euler equation:
\begin{equation}
\underbrace{\left[ 
\begin{array}{c}
\mathbf{F}_{\mathrm{ext}/\mathcal{S},G}\\
\mathbf{T}_{\mathrm{ext}/\mathcal{S},G}
\end{array}\right]}_{\mathbf{W}_{\mathrm{ext}/\mathcal{S},G}}+
\underbrace{\left[ 
\begin{array}{c}
\mathbf{F}_{\mathcal{A}/\mathcal{S},G}\\
\mathbf{T}_{\mathcal{A}/\mathcal{S},G}
\end{array}\right]}_{\mathbf{W}_{\mathcal{A}/\mathcal{S},G}}=
\underbrace{
\left[\begin{array}{cc}
m^{\mathcal{S}}\mathbf{I}_3 & \mathbf{0}_{3\times 3} \\
\mathbf{0}_{3\times 3} & \mathbf{I}_G^{\mathcal{S}}
\end{array}
\right]
}_{\mathcal{M}_G^{\mathcal{S}}}
\underbrace{
\left[\begin{array}{c}
\ddot{\mathbf{a}}_G \\
\dot{\bm{\omega}}_G
\end{array}
\right]
}_{\ddot{\mathbf{q}}_G}+
\left[
\begin{array}{c}
\mathbf{0}_{3\times3} \\
\left[\bm{\omega_G}\right]_{\times} \mathbf{I}^{\mathcal{S}}_{G}\bm{\omega_G}
\end{array}
\right],
\label{eq:dynamics_G}
\end{equation}
where $\mathbf{W}_{\mathcal{A}/\mathcal{S},G}$ is the wrench of the forces/torques applied by the appendage $\mathcal{A}$ to the main hub referred to the point $G$, $m^{\mathcal{S}}$ is the mass of the main hub, $\mathbf{I}_{G}^{\mathcal{S}}$ is the inertia matrix of the main hub w.r.t. $G$ and $\left[\bm{\omega_G}\right]_{\times}$ is the skew matrix associated to the angular velocity $\bm{\omega}_{G}$. associated to the vector $\mathbf{r}_{GC}$ from node $G$ to node $C$.

The wrench $\mathbf{W}_{\mathcal{A}/\mathcal{S},G}$ and $\mathbf{W}_{\mathcal{A}/\mathcal{S},C}$ are connected by the the \textit{kinematic model} (or \textit{Jacobian}) $\bm{\tau}_{CG}$:

\begin{equation}
\underbrace{\left[
\begin{array}{c}
\mathbf{F}_{\mathcal{A}/\mathcal{S},G} \\
\mathbf{T}_{\mathcal{A}/\mathcal{S},G}
\end{array}
\right]}_{\mathbf{W}_{\mathcal{A}/\mathcal{S},G}} =
\underbrace{\left[
\begin{array}{cc}
\mathbf{I}_3 & \mathbf{0}_{3\times 3} \\
-\left[\mathbf{r}_{GC}\right]_{\times} & \mathbf{I}_3
\end{array}
\right]}_{\bm{\tau}_{CG}^{\mathrm{T}}}
\underbrace{\left[
\begin{array}{c}
\mathbf{F}_{\mathcal{A}/\mathcal{S},C} \\
\mathbf{T}_{\mathcal{A}/\mathcal{S},C}
\end{array}
\right]}_{\mathbf{W}_{\mathcal{A}/\mathcal{S},C}}.
\label{eq:wrench_G}
\end{equation}

If now we consider the point $C$ on the body $\mathcal{S}$, the translation velocity at the point $C$ can be expressed as a function of the velocity $\mathbf{v}_G$ at point $G$ and the angular velocity $\bm{\omega}_G$:
\begin{equation}
\mathbf{v}_C = \mathbf{v}_G + \left[\bm{\omega}_G\right]_{\times}\mathbf{r}_{GC} = \mathbf{v}_G - \left[\mathbf{r}_{GC}\right]_{\times}\bm{\omega}_G.
\label{eq:velocity_c0}
\end{equation}
The complete velocity vector at point C is written as:
\begin{equation}
\left[
\begin{array}{c}
\mathbf{v}_C \\
\bm{\omega}_C
\end{array}
\right] = \underbrace{
\left[
\begin{array}{cc}
\mathbf{I}_3 & -\left[\mathbf{r}_{GC}\right]_{\times} \\
\mathbf{0}_{3\times 3} & \mathbf{I}_3
\end{array}
\right]
}_{\bm{\tau}_{CG}}
\left[
\begin{array}{c}
\mathbf{v}_{G} \\
\bm{\omega}_G
\end{array}
\right],
\label{eq:velocity_c}
\end{equation}
The time-derivation of Eq. (\ref{eq:velocity_c0}) in the inertial frame gives:
\begin{equation}
\mathbf{a}_C = \mathbf{a}_{G} - \left[\left.\frac{d\mathbf{r}_{GC}}{dt}\right\vert_{\mathcal{S}}\right]_{\times}\bm{\omega}_G - \left[\bm{\omega}_G\right]_{\times}\left[\mathbf{r}_{GC}\right]_{\times}\bm{\omega}_G.
\end{equation}
Since the body $\mathcal{S}$ is considered rigid by assumption, the derivative $\left.\frac{d\mathbf{r}_{GC}}{dt}\right\vert_{\mathcal{S}}=0$ and
\begin{equation}
\underbrace{\left[
\begin{array}{c}
\mathbf{a}_C \\
\dot{\bm{\omega}}_C
\end{array}
\right]}_{\ddot{\mathbf{q}}_C}=
\underbrace{
\left[
\begin{array}{cc}
\mathbf{I}_3 & -\left[ \mathbf{r}_{GC}\right]_{\times} \\
\mathbf{0}_{3\times 3} & \mathbf{I}_3
\end{array}
\right]}_{\bm{\tau}_{CG}}
\underbrace{\left[\begin{array}{c}
\mathbf{a}_G \\
\dot{\bm{\omega}}_G
\end{array}\right]}_{\ddot{\mathbf{q}}_G}+
\left[
\begin{array}{c}
- \left[\bm{\omega}_G\right]_{\times}\left[\mathbf{r}_{GC}\right]_{\times}\bm{\omega}_G \\
\mathbf{0}_{3\times 1}
\end{array}
\right].
\label{eq:a_c}
\end{equation}

Generally, for spacecraft in operation mode, the angular velocity $\bm{\omega}_G$ is very small. Thanks to this assumption it is possible to linearize the Eqs. \eqref{eq:dynamics_G} and \eqref{eq:a_c}, where all the quadratic terms $\left[\bm{\omega}_G\right]_{\times}\mathbf{X}_{3\times 3}\bm{\omega}_G$ can be neglected. 
By using Eq. \eqref{eq:wrench_G} we rewrite Eqs. \eqref{eq:dynamics_G} and \eqref{eq:a_c}:
\begin{equation}
\mathbf{W}_{\mathrm{ext}/\mathcal{S},G}+\bm{\tau}_{CG}^{\mathrm{T}}\mathbf{W}_{\mathcal{A}/\mathcal{S},C}=\mathcal{M}_G^{\mathcal{S}}\ddot{\mathbf{q}}_G,
\label{eq:titop_sc1}
\end{equation}
\begin{equation}
\ddot{\mathbf{q}}_{C} = \bm{\tau}_{CG}\ddot{\mathbf{q}}_{G}.
\label{eq:titop_sc2}
\end{equation}

From Eqs. \eqref{eq:titop_sc1} and \eqref{eq:titop_sc2} the TITOP model of the spacecraft is directly obtained:
\begin{equation}
\left[
\begin{array}{c}
\ddot{\mathbf{q}}_C \\
\ddot{\mathbf{q}}_G
\end{array}
\right]=
\underbrace{
\left[
\begin{array}{c}
\bm{\tau}_{CG} \\
\mathbf{I}_6
\end{array}
\right]
\left[
\mathcal{M}_G^{\mathcal{S}}
\right]^{-1}
\left[
\begin{array}{cc}
\bm{\tau}_{CG}^{\mathrm{T}} & \mathbf{I}_{6}
\end{array}
\right]
}_{\left[
\tilde{\mathcal{M}}_{GC}^{\mathcal{S}}
\right]^{-1}
}
\left[
\begin{array}{c}
\mathbf{W}_{\mathcal{A}/\mathcal{S},C} \\
\mathbf{W}_{\mathrm{ext}/\mathcal{S},G}
\end{array}
\right].
\end{equation}

This model can be easily extended to the case of $n_{\mathcal{A}}$ appendages attached to the main hub at the points $C_1$, $C_2$, $\dots$, $C_{n_{\mathcal{A}}}$:

\begin{equation}
\left[
\begin{array}{c}
\ddot{\mathbf{q}}_{C_1} \\
\ddot{\mathbf{q}}_{C_2} \\
\vdots \\
\ddot{\mathbf{q}}_{C_{n}} \\
\ddot{\mathbf{q}}_G
\end{array}
\right]=
\underbrace{
\left[
\begin{array}{c}
\bm{\tau}_{C_1 G} \\
\bm{\tau}_{C_2 G} \\
\vdots \\
\bm{\tau}_{C_{n} G} \\
\mathbf{I}_6
\end{array}
\right]
\left[
\mathcal{M}_G^{\mathcal{S}}
\right]^{-1}
\left[
\begin{array}{ccccc}
\bm{\tau}_{C_1 G}^{\mathrm{T}} & \bm{\tau}_{C_2 G}^{\mathrm{T}} & \dots & \bm{\tau}_{C_{n} G}^{\mathrm{T}} & \mathbf{I}_{6}
\end{array}
\right]
}_{\left[
\tilde{\mathcal{M}}_{GC_1\dots C_n}^{\mathcal{S}}
\right]^{-1}
}
\left[
\begin{array}{c}
\mathbf{W}_{\mathcal{A}_1/\mathcal{S},C_1} \\
\mathbf{W}_{\mathcal{A}_2/\mathcal{S},C_2} \\
\vdots \\
\mathbf{W}_{\mathcal{A}_{n_{\mathcal{A}}}/\mathcal{S},C_{n}} \\
\mathbf{W}_{\mathrm{ext}/\mathcal{S},G}
\end{array}
\right].
\label{eq:spacecraft_nappendages}
\end{equation}

The block diagram of the model in Eq. \eqref{eq:spacecraft_nappendages} is shown in Fig. \ref{fig:satellite_titop}.

\subsection{Generic flexible appendage model}
\label{sec:model_appendage}

Let consider a simple flexible appendage $\mathcal{A}$ attached at the point $P$ of the main hub $\mathcal{S}$. The dynamic model between the acceleration vector $\ddot{\mathbf{q}}_P$ of the point $P$ and the wrench $\mathbf{W}_{\mathcal{A}/\mathcal{S},P}$ of the forces/torques imposed by $\mathcal{A}$ to $\mathcal{S}$ is:

\begin{equation}
\left[
\begin{array}{c}
\dot{\bm{\eta}}\\
\ddot{\bm{\eta}} \\ \hline
\mathbf{W}_{\mathcal{A}/\mathcal{S},P}
\end{array}
\right]= 
\underbrace{\left[
\begin{array}{cc|c}
\mathbf{0}_{N\times N} & \mathbf{I}_N & \mathbf{0}_{N\times 6} \\
-\mathrm{diag}(\omega_i^2) & -\mathrm{diag}(2\zeta_i\omega_i) & -\mathbf{L}_P \\ \hline
-\mathbf{L}_P^{\mathrm{T}}\mathrm{diag}(\omega_i^2) & -\mathbf{L}_P^{\mathrm{T}}\mathrm{diag}(2\zeta_i\omega_i) & \mathcal{D}_{P_0}^{\mathcal{A}}
\end{array}
\right]}_{\mathcal{M}_{P}^{\mathcal{A}}}
\left[
\begin{array}{c}
\bm{\eta} \\
\dot{\bm{\eta}} \\ \hline
\ddot{\mathbf{q}}_P
\end{array}
\right],
\label{eq:appendage_titop}
\end{equation}

where $N$ is the number of the considered flexible modes, $\omega_i$, $\eta_i$ are respectively the $i$-th flexible mode's angular frequency and damping ratio, $\mathbf{L}_P$ is the matrix $N\times 6$ of the modal participation factors expressed at point $P$, $\mathcal{D}_{P_0}^{\mathcal{A}} = \mathcal{D}_P^{\mathcal{A}} - \mathbf{L}_P^{\mathrm{T}}\mathbf{L}_P$ is the so called \textit{residual mass matrix}, where $\mathcal{D}_{P}^{\mathcal{A}}$ is the static model of the appendage w.r.t. the attachment point $P$:
\begin{equation}
\mathcal{D}^{\mathcal{A}}_P = \bm{\tau}_{AP}^{\mathrm{T}}\left[
\begin{array}{cc}
m^{\mathcal{A}}\mathbf{I}_3 & \mathbf{0}_{3\times 3} \\
\mathbf{0}_{3\times 3} & \mathbf{I}^{\mathcal{A}}_{A} 
\end{array}
\right]\bm{\tau}_{AP},
\end{equation}
where $m^{\mathcal{A}}$ is the mass of the appendage $\mathcal{A}$ and $\mathbf{I}^{\mathcal{A}}_{A}$ is the inertia matrix of the appendage $\mathcal{A}$ expressed at the appendage center of mass $A$. 

\begin{figure*}[tb]
	\centering
	\begin{minipage}[b]{.34\linewidth}
		\centering
		$\vcenter{\hbox{\includegraphics[width=\columnwidth]{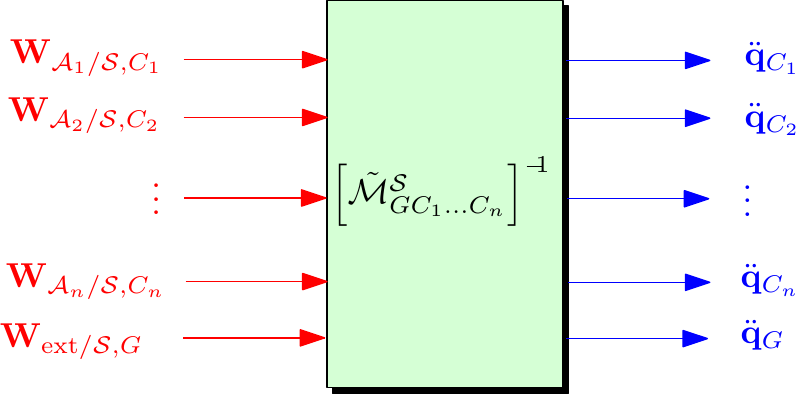}}}$
	\end{minipage}%
	\hfill%
	\begin{minipage}[b]{.63\linewidth}
		\centering
		$\vcenter{\hbox{\includegraphics[width=.9\columnwidth]{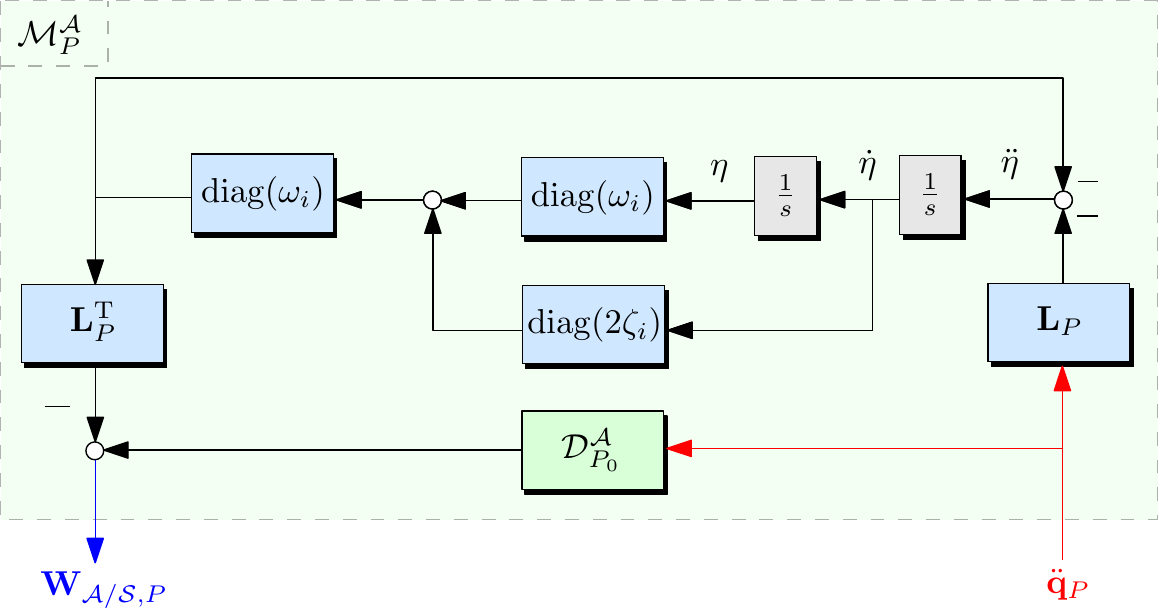}}}$
	\end{minipage}\\[2pt]
	\begin{minipage}[t]{.34\linewidth}
		\caption{Spacecraft body diagram.}
		\label{fig:satellite_titop}
	\end{minipage}%
	\hfill%
	\begin{minipage}[t]{.63\linewidth}
		\caption{Single port model of a flexible appendage clamped at point $P$.}
		\label{fig:appendage_titop}
	\end{minipage}%
\end{figure*}

The block diagram equivalent to Eq. \eqref{eq:appendage_titop} with the minimum number of occurrences of $\omega_i$, $\eta_i$ and $\mathbf{L}_P$ is shown in Fig. \ref{fig:appendage_titop}.

\subsection{Connection between central body and flexible appendage}
\label{sec:connection}
\begin{figure*}[ht]
	\centering
	\begin{minipage}[b]{.34\linewidth}
		\centering
		$\vcenter{\hbox{\includegraphics[width=\columnwidth]{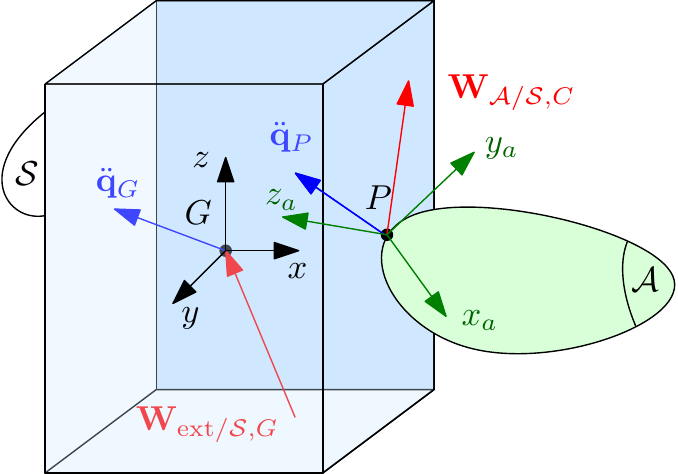}}}$
	\end{minipage}%
	\hfill%
	\begin{minipage}[b]{.63\linewidth}
		\centering
		$\vcenter{\hbox{\includegraphics[width=\columnwidth]{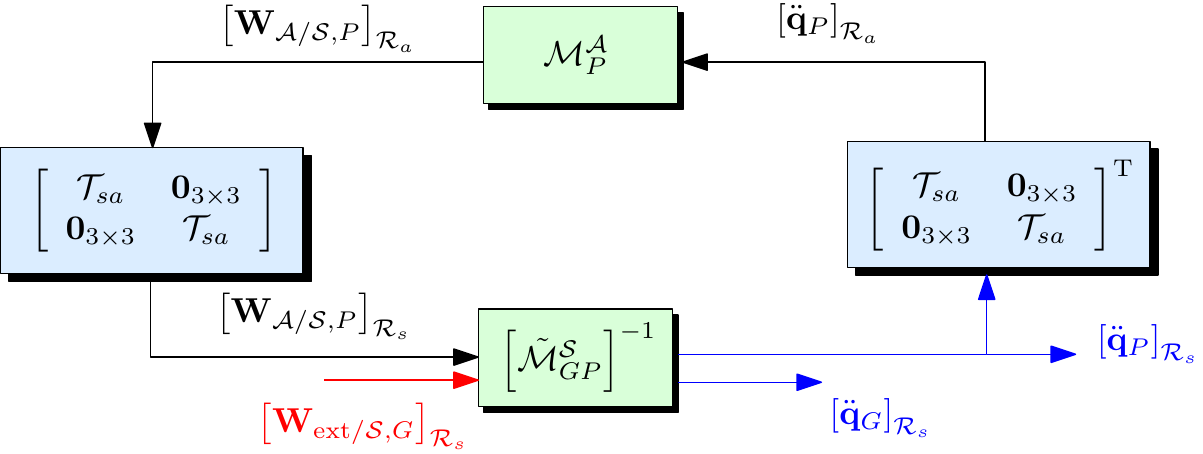}}}$
	\end{minipage}\\[2pt]
	\begin{minipage}[t]{.34\linewidth}
		\caption{Connection between the main hub and a flexible appendage.}
		\label{fig:satAndAppengage}
	\end{minipage}%
	\hfill%
	\begin{minipage}[t]{.63\linewidth}
		\caption{TITOP model of a spacecraft with a flexible appendage.}
		\label{fig:titop_sat_append}
	\end{minipage}%
\end{figure*}

Let consider Fig. \ref{fig:satAndAppengage} where a flexible appendage $\mathcal{A}$ is connected to the spacecraft $\mathcal{S}$. We found the model of the rigid spacecraft in Eq. (\ref{eq:spacecraft_nappendages}) expressed in the spacecraft frame $\mathcal{R}_{s}=(G;\mathbf{x},\mathbf{y},\mathbf{z})$ and the model of a flexible appendage in Eq. (\ref{eq:appendage_titop}) expressed in the appendage frame $\mathcal{R}_a=(P;\mathbf{x}_a,\mathbf{y}_a,\mathbf{z}_a)$. A cosine direction matrix $\mathcal{T}_{sa}$ of the rotation from frame $\mathcal{R}_a$ to frame $\mathcal{R}_s$ is thus needed to connect the two models. $\mathcal{T}_{sa}$ contains the coordinates of vectors $\mathbf{x}_a$, $\mathbf{y}_a$, $\mathbf{z}_a$ expressed in frame $\mathcal{R}_s$. Figure \ref{fig:titop_sat_append} shows how $\mathcal{A}$ is connected to $\mathcal{S}$. The matrix $\mathcal{T}_{sa}$ has a constant value if the appendage is fixed in a certain position to the main hub. On the other hand if the appendage rotates around a direction $\mathbf{r}_a$, $\mathcal{T}_{sa}$ is a function of the rotation angle $\alpha$. This is for example the case of rotating solar panels. 

Let consider in Fig. \ref{fig:satAndAppengage} the simple case where the appendage's axis $z_a$ is aligned with the spacecraft axis $z$ and that $\mathcal{A}$ rotates around $z_a$.  
In this case $\mathcal{T}_{sa}$ takes the form:
\begin{equation}
	\mathcal{T}_{sa}(\alpha) = \left[
	\begin{array}{ccc}
		\cos\alpha & -\sin\alpha & 0 \\
		\sin\alpha & \cos\alpha & 0 \\
		0 & 0 & 1
	\end{array}
	\right].
	\label{eq:t_sa}
\end{equation} 

The matrix in Eq. (\ref{eq:t_sa}) can be put in LFT form as demonstrated by Guy \textit{et al.} \cite{Guy2014}. Since the trigonometrical functions $\cos$ and $\sin$ are not rational expressions in $\alpha$, a new parameter $\tau=\tan(\alpha/4)$ is introduced.

Thanks to this model it is possible to obtain all the dynamic behaviors of the spacecraft on a whole rotation ($\tau\in\left]-1;1\right]$) with only 4 occurrences of the block $\tau$ as demonstrated by \cite{dubanchet2016modeling}. Since four blocks $\mathcal{T}_{sa}$ are required for the connection of the appendage on the main hub, the complete model presented in Fig. \ref{fig:titop_sat_append} will result in 16 occurrences of $\tau$.

\subsection{Solar Array Drive Mechanism dynamic model}
    \subsubsection{Stepper motor model}\label{sec:step_motor}
    
    An SADM is generally composed of three elements: a stepper motor, an electronic driver and a reduction gearbox (when foreseen).
The stepper motors classically used in space systems are bi-phase permanent magnet-like. This type of motors employs permanent magnets connected to the rotor axis and bi-phased wires alternately disposed and regularly spaced on the stator. The rotor, composed of $p$ magnetic poles, is steered by the interaction between the magnetic field produced by the pair of poles and the stator wires. 

The dynamics of the stepper motor is described by the model \cite{Chen2016}:
\begin{equation}
J_r\ddot{\theta}_r = T_e - C_r\dot{\theta}_r - T_l,
\label{eq:sadm_eq}
\end{equation}

where $J_r$ is the moment of the rotor inertia, $\theta_r$ is the angular displacement of the rotor, $T_e$ is the electromagnetic torque, $C_r$ is the viscous damping coefficient and $T_l$ is the load torque.
The electromagnetic torque can be then expressed as:
\begin{equation}
T_e = \underbrace{K_m\left[I_B\cos(z\theta_r)-I_A\sin(z\theta_r)\right]}_{T_m}-\underbrace{K_d\sin(4z\theta_r)}_{T_d},
\label{eq:te_motor}
\end{equation}
where $I_A$ and $I_B$ are respectively the currents in the phases A and B, $z$ is the number of rotor teeth, and $K_m$ and $K_d$ are respectively the motor and detent torque constants. The quantities $T_m$ and $T_d$ are respectively defined as \textit{motor} and \textit{detent torque}.

The motor constant $K_m$ depends on the motor magnetic flux and the number of stator teeth $z$ \cite{Chen2016}.

The \textit{detent torque} is a passive electromagnetic torque created by the interaction between the rotor permanent magnets and the stator wires even in absence of current in the same wires. Even if in reality the detent torque's shape varies from one step to the other, $T_d$ can be modeled as an harmonic signal with spatial period correspondent to a step. Since the number of steps corresponds to the quadruple of the number of rotor teeth $z$, $T_d$ is expressed as defined in Eq. \eqref{eq:te_motor}.
The coefficient $K_d$ depends on the physical characteristic of the motor as the quantity of ferromagnetic material at the stator level.

Let consider for the moment that the detent torque is negligible as for most of the real applications where $T_d \ll T_m$.
Equation (\ref{eq:te_motor}) becomes:
\begin{equation}
T_e = K_m\left[I_B\cos(z\theta_r)-I_A\sin(z\theta_r)\right],
\label{eq:te_motor2}
\end{equation}
The current in the wires provided by the electronic driver is a stair step sinusoidal curve. In order to improve the smoothness of the rotor angular displacement a sub-division in $n_\mu$ micro-steps of the full electrical step $2\pi/p$ is generally needed. After subdivision the expression of $I_A$ and $I_B$ results:
\begin{equation}
I_A = I\cos(\gamma i), \quad I_B = I\sin(\gamma i),
\label{eq:IaIb}
\end{equation}
where $I$ is the maximum value of the two-phase current, $\gamma = 2\pi/(p n_\mu)$ is the electrical $\mu$-step angle after subdivision $n_\mu$  and $i=1,2,\dots$ is the step number. By substitution of Eq. \eqref{eq:IaIb} in Eq. \eqref{eq:te_motor2} the electromagnetic torque results:
\begin{equation}
T_e = K_mI\sin(\gamma i-z\theta_r)=K_mI\sin\left[z\left(\frac{\gamma i}{z}-\theta_r\right)\right]= K_mI\sin\left(z\Delta\theta_r\right),
\label{eq:te_nonlin}
\end{equation}
where $\Delta\theta_r$ represents the mechanical angle between the actual position on the rotor and the theoretical equilibrium position between two consecutive pulse signals. The condition to be satisfied not to lose any step is \cite{Chen2016}:
\begin{equation}
-\alpha \leq \Delta\theta_r \leq \alpha,
\end{equation}
where $\alpha = 2\pi/\left(zpn_\mu\right)$ is the mechanical $\mu$-step angle after subdivision. As consequence:
\begin{equation}
-\frac{2\pi}{pn_\mu} \leq z\Delta\theta_r \leq \frac{2\pi}{pn_\mu}
\end{equation}
This condition, which constraints the angle $z\Delta\theta_r$ to small values for typical values of $p$ and $n_\mu$, allows us to linearize Eq. \eqref{eq:te_nonlin}:
\begin{equation}
T_e = K_mI\sin\left(z\Delta\theta_r\right)\approx K_mI\gamma i - K_mIz\theta_r.
\label{eq:Te_lin}
\end{equation}
Substituting Eq. (\ref{eq:Te_lin}) into Eq. (\ref{eq:sadm_eq}), we find:
\begin{equation}
J_r\ddot{\theta}_r + C_r\dot{\theta}_r + K_0\theta_r = K_mI\gamma i - T_l,
\label{eq:sadm_mck}
\end{equation}
with
\begin{equation}
K_0 = K_m I z.
\end{equation}
Equation \eqref{eq:sadm_mck} corresponds to the dynamics of a classical mechanical system with stiffness $K_0$ (that is defined in this case as \textit{electromagnetic stiffness}) and exciting torque $K_mI\gamma i - T_l$. However this equation presents two limitations: only the dynamics around the rotor axis is taken into account; the equation is valid for a SADM fixed to an inertial frame.    
The model \eqref{eq:sadm_mck} can be generalized to the case where the SADM is mounted on a free platform as a spacecraft. 

\begin{figure}[tb]
\centering
\includegraphics[width=0.7\columnwidth]{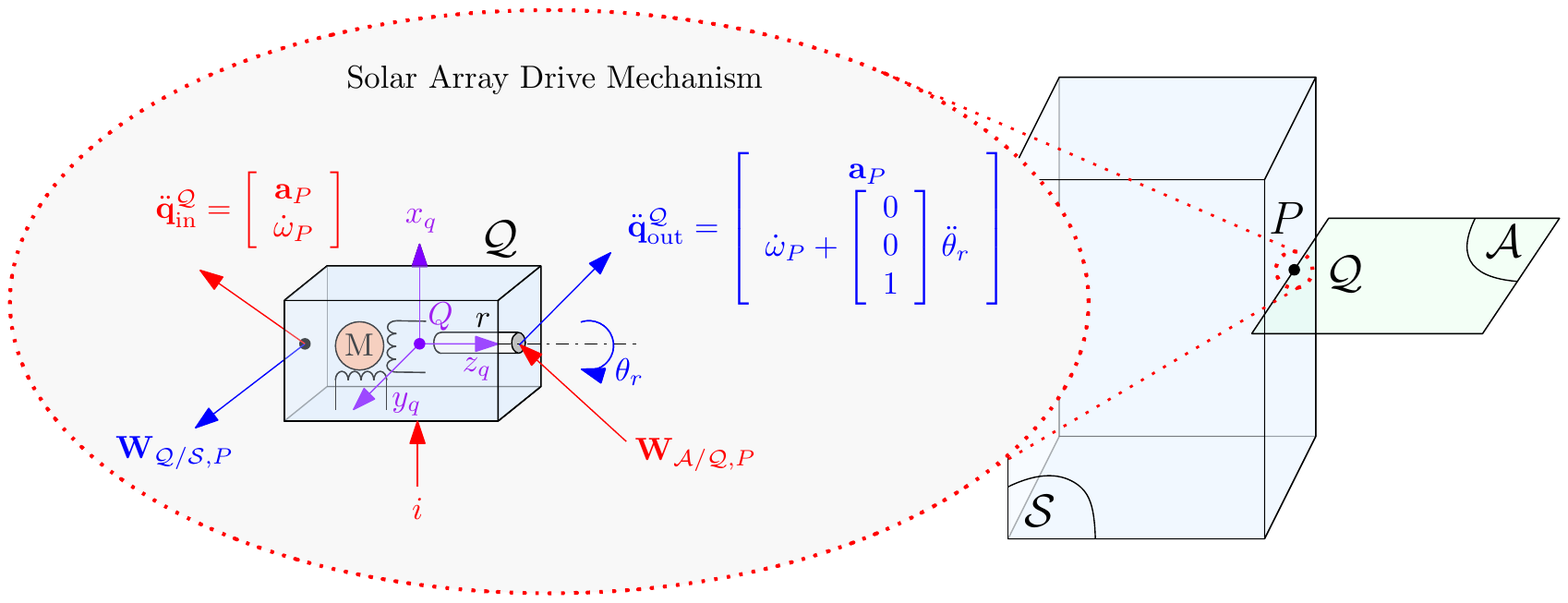} 
\caption{Stepper motor TITOP scheme with revolute joint axis $\mathbf{z}_q$ aligned with $\mathbf{z}_a$.} 
\label{fig:sadm_titop_scheme}
\end{figure}

Let us consider Fig. \ref{fig:sadm_titop_scheme}. The stepper motor $\mathcal{Q}$ is considered as a massless intermediate body between the main hub $\mathcal{S}$ and the solar panel $\mathcal{A}$, centered at point $Q$ (coincident with $P$). As shown in the figure the SADM can be modeled as an augmented (a third port is added to take into account the command $i$ and the rotor angle $\theta_r$) TITOP system written in the appendage frame $\mathcal{R}_a = \left(P;\mathbf{x}_a,\mathbf{y}_a,\mathbf{z}_a\right)$, where the inputs are:
\begin{itemize}[noitemsep,topsep=0pt]
\item the wrench $\mathbf{W}_{\mathcal{A}/\mathcal{Q},P} = \left[\mathbf{F}_{\mathcal{A}/\mathcal{Q},P}^{\mathrm{T}}\,\mathbf{T}_{\mathcal{A}/\mathcal{Q},P}^{\mathrm{T}} \right]^{\mathrm{T}} \in \mathbb{R}^{6\times 1}$ of forces/torques imposed by the appendage $\mathcal{A}$ to the SADM $\mathcal{Q}$ at the point $P$,
\item the acceleration vector $\ddot{\mathbf{q}}_{\mathrm{in}}^{\mathcal{Q}} \in \mathbb{R}^{6\times 1}$ of the SADM casing, 
\item the step number $i$,
\end{itemize}
and the outputs are:
\begin{itemize}[noitemsep,topsep=0pt]
\item the acceleration vector $\ddot{\mathbf{q}}_{\mathrm{out}}^{\mathcal{Q}}\in \mathbb{R}^{6\times 1}$ of the SADM where the rotor acceleration $\ddot{\theta}_r$ is added to the third component of the angular accelerations (along $\mathbf{z}_q$). This hypothesis is not restrictive and can be generalized for any rotating axis. In case the revolute joint axis is along an arbitrary unit axis $\mathbf{r}_{a}=\left[x_{r_a}\,y_{r_a}\,z_{r_a}\right]^{\mathrm{T}}_{\mathcal{R}_a}$, the direction cosine matrix:
\begin{equation}
\mathbf{P}_{r/a} = \mathbf{P}_{\mathbf{r}_a}\,\mathrm{diag}\left(\mathrm{det}\left(\mathbf{P}_{\mathbf{r}_a}\right),\,1,\,1\right)    
\end{equation}
with $\mathbf{P}_{\mathbf{r}_a} = \left[\mathrm{ker}\left(\left[\mathbf{r}_a\right]^{\mathrm{T}}_{\mathcal{R}_a}\right)\quad \left[\mathbf{r}_a\right]_{\mathcal{R}_a}\right]$
allows the TITOP model $\mathcal{M}_P^{\mathcal{Q}}$ to be generalized. In the following paragraph we consider that the revolute joint axis $\mathbf{z}_q$ is aligned with $\mathbf{z}_a$ to make the equations and the block diagrams more readable. Reference frames $\mathcal{R}_a$ and $\mathcal{R}_q$ coincide in this particular case. 
\item the wrench $\mathbf{W}_{\mathcal{Q}/\mathcal{S},P}= \left[\mathbf{F}_{\mathcal{Q}/\mathcal{S},P}^{\mathrm{T}}\,\mathbf{T}_{\mathcal{Q}/\mathcal{S},P}^{\mathrm{T}} \right]^{\mathrm{T}} \in \mathbb{R}^{6\times 1}$ of forces/torques imposed by the SADM $\mathcal{Q}$ to the main hub $\mathcal{S}$ at point $P$,
\item the rotor angle $\theta_r$.
\end{itemize}

Some observations help to develop the TITOP model of the SADM:
\begin{itemize}[noitemsep,topsep=0pt]
\item the angular acceleration $\dot{\bm{\omega}}^{\mathcal{Q}}_\mathrm{in}$ of the SADM casing expressed in $\mathcal{R}_S$ is equal to the angular acceleration $\dot{\bm{\omega}}_P$ at the attachment point $P$,
\item the angular acceleration at SADM output $\dot{\bm{\omega}}_\mathrm{out}^{\mathcal{Q}}$ expressed in $\mathcal{R}_q = \left(Q;\mathbf{x}_q,\mathbf{y}_q,\mathbf{z}_q\right)$ is equal to the angular acceleration $\dot{\bm{\omega}}_P$ at the attachment point $P$ plus the contribution of the rotor acceleration $\ddot{\theta}_r$: $\dot{\bm{\omega}}_{\mathrm{out}}^{\mathcal{Q}}=\dot{\bm{\omega}}_P+\left[\begin{array}{ccc}
0 & 0 & 1
\end{array}\right]^{\mathrm{T}}\ddot{{\theta}}_r$,
\item the force vector imposed by the appendage to the SADM is directly transmitted to the spacecraft: $\mathbf{F}_{\mathcal{Q}/\mathcal{S}}=\mathbf{F}_{\mathcal{A}/\mathcal{Q}}$,
\item the first two components of the torque imposed by the appendage to the SADM are directly transmitted to the spacecraft: $\mathbf{T}_{\mathcal{Q}/\mathcal{S},P}(1:2) =\mathbf{T}_{\mathcal{A}/\mathcal{Q},P}(1:2)$,
\item the third component of the torque transmitted by the SADM to the spacecraft structure is exactly equal and opposite to the torque developed by the stepper motor: $\mathbf{T}_{\mathcal{Q}/\mathcal{S},P}(3)=-\left(K_m I\gamma i - C_r\dot{\theta}_r - K_0\theta_r\right) = -C_m$,
\item by taking into account the relative motion of the rotor w.r.t. the SADM casing, Eq. \eqref{eq:sadm_mck} becomes:
\begin{equation}
J_r\left(\ddot{\theta}_r + \dot{\bm{\omega}}_{\mathrm{in}}^{\mathcal{Q}}(3)\right) + C_r\dot{\theta}_r + K_0\theta_r = K_m I\gamma i - T_l
\label{eq:sadm_eq}
\end{equation} 
If the revolute joint axis is along an arbitrary unit axis $\mathbf{r}_{a}=\left[x_{r_a}\,y_{r_a}\,z_{r_a}\right]^{\mathrm{T}}_{\mathcal{R}_a}$ Eq. \eqref{eq:sadm_eq} becomes:
\begin{equation}
    J_r\left(\ddot{\theta}_r +\left[\mathbf{r}_a\right]^{\mathrm{T}}_{\mathcal{R}_a} \left[\dot{\bm{\omega}}_{\mathrm{in}}^{\mathcal{Q}}\right]_{\mathcal{R}_a}\right) + C_r\dot{\theta}_r + K_0\theta_r = K_m I\gamma i - T_l
\end{equation}
\item the load torque $T_l$ in (\ref{eq:sadm_mck}) is equal and opposite to the torque imposed by the appendage to the SADM: $T_l = - \mathbf{T}_{\mathcal{A}/\mathcal{Q},P}(3)$ along the mechanism axis if $\mathbf{z}_q\equiv\mathbf{z}_a$. If the revolute joint axis is along an arbitrary unit axis $\mathbf{r}_{a}=\left[x_{r_a}\,y_{r_a}\,z_{r_a}\right]^{\mathrm{T}}_{\mathcal{R}_a}$, $T_l = -\left[\mathbf{r}_a\right]^{\mathrm{T}}_{\mathcal{R}_a}\left[\mathbf{T}_{\mathcal{A}/\mathcal{Q},P}\right]_{\mathcal{R}_a}=-\left[\mathbf{T}_{\mathcal{A}/\mathcal{Q},P}\right]_{\mathcal{R}_q}(3)$.
\end{itemize}

\begin{figure}[tb]
\centering
\includegraphics[width=.8\columnwidth]{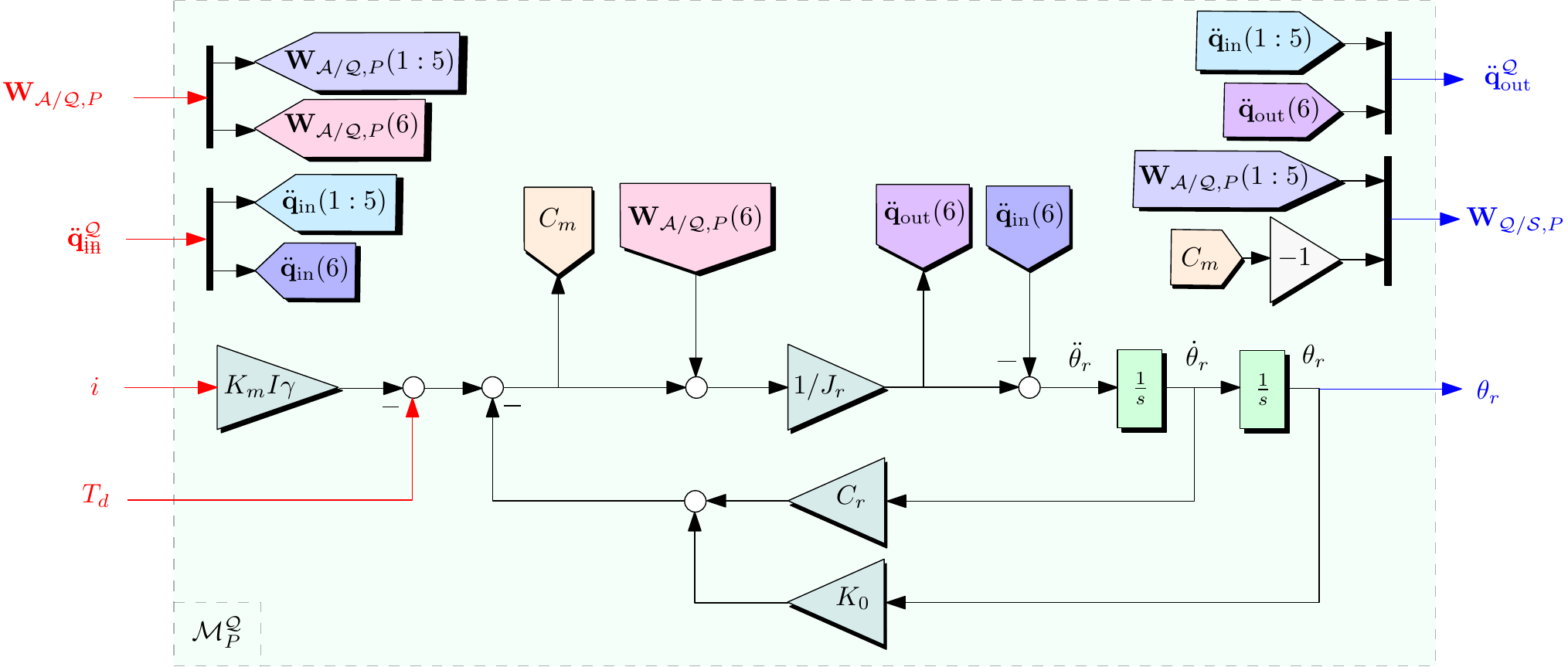} 
\caption{SADM TITOP block diagram with revolute joint axis $\mathbf{z}_q$ aligned with $\mathbf{z}_a$.} 
\label{fig:sadm_titop_block_diagram}
\end{figure}

Using the previous assumptions it is possible to write the augmented TITOP model $\mathcal{M}_P^{\mathcal{Q}}$ of the SADM ($\mathbf{z}_q\equiv\mathbf{z}_a$) as done in Eq. \eqref{eq:sadm_titop}, whose block diagram is shown in Fig. \ref{fig:sadm_titop_block_diagram}.

\subsubsection{Reduction gearbox model}
\label{sec:model_gearbox}
The connection between the main hub and a solar array through a SADM system can be direct or driven by a reduction gearbox $\mathcal{G}$. The gearbox model depends on the following parameters: the gearbox ratio $N_g$, the stiffness $K_g$, the damping $C_g$, the inertia of the output shaft $J_o$, the inertia of the input shaft $J_i$, the direction $\mathbf{r}_a$ of the gear axis in load (solar array) frame $\mathcal{R}_a$.

Let define:
\begin{itemize}[noitemsep,topsep=0pt]
    \item $T_{/o} = \left[\mathbf{r}_a\right]^{\mathrm{T}}_{\mathcal{R}_a}\left[\mathbf{T}_{\mathcal{A}/\mathcal{G},P}\right]_{\mathcal{R}_a}$ the torque applied by the solar array on the output shaft of the gearbox at the connection point $P$ around $\mathbf{r}_a$;
    \item $T_{i/} = \left[\mathbf{r}_a\right]^{\mathrm{T}}_{\mathcal{R}_a}\left[\mathbf{T}_{\mathcal{G}/\mathcal{Q},P}\right]_{\mathcal{R}_a}$ the torque applied by the input shaft of the gearbox to the stepper motor rotor at the connection point $P$ around $\mathbf{r}_a$;
    \item $\ddot{\theta}_o = \left[\mathbf{r}_a\right]^{\mathrm{T}}_{\mathcal{R}_a}\left[\dot{\bm{\omega}}_\mathrm{out}^{\mathcal{G}}\right]_{\mathcal{R}_a}$ the angular acceleration of the output shaft around $\mathbf{r}_a$;
    \item $\ddot{\theta}_i = \left[\mathbf{r}_a\right]^{\mathrm{T}}_{\mathcal{R}_a}\left[\dot{\bm{\omega}}_\mathrm{in}^{\mathcal{G}}\right]_{\mathcal{R}_a}$ the angular acceleration of the input shaft around $\mathbf{r}_a$.
\end{itemize}
The angular deflection of the gearbox due to its stiffness, seen from the output shaft is:
\begin{equation}
    \delta \theta_o = \theta_o - \frac{\theta_i}{N_g}
\end{equation}
and the internal reaction torque on the output shaft can be described by:
\begin{equation}
    T_{\mathcal{G}/o} = -K_g\delta\theta_o - C_g\delta\dot{\theta}_o + T_{gb},
\end{equation}
where $T_{gb}$ is an harmonic disturbance due to one or more contact imperfection frequencies in the gear pairs inside the gearbox.  
The dynamic model of the gearbox then reads:
\begin{align}
    J_o\ddot{\theta}_o &= T_{/o} + T_{\mathcal{G}/o} \\
    J_i\ddot{\theta}_i &= -T_{i/}-T_{\mathcal{G}/o}/N_g
\end{align}

\begin{figure}[tb]
	\centering
	\includegraphics[width=.8\columnwidth]{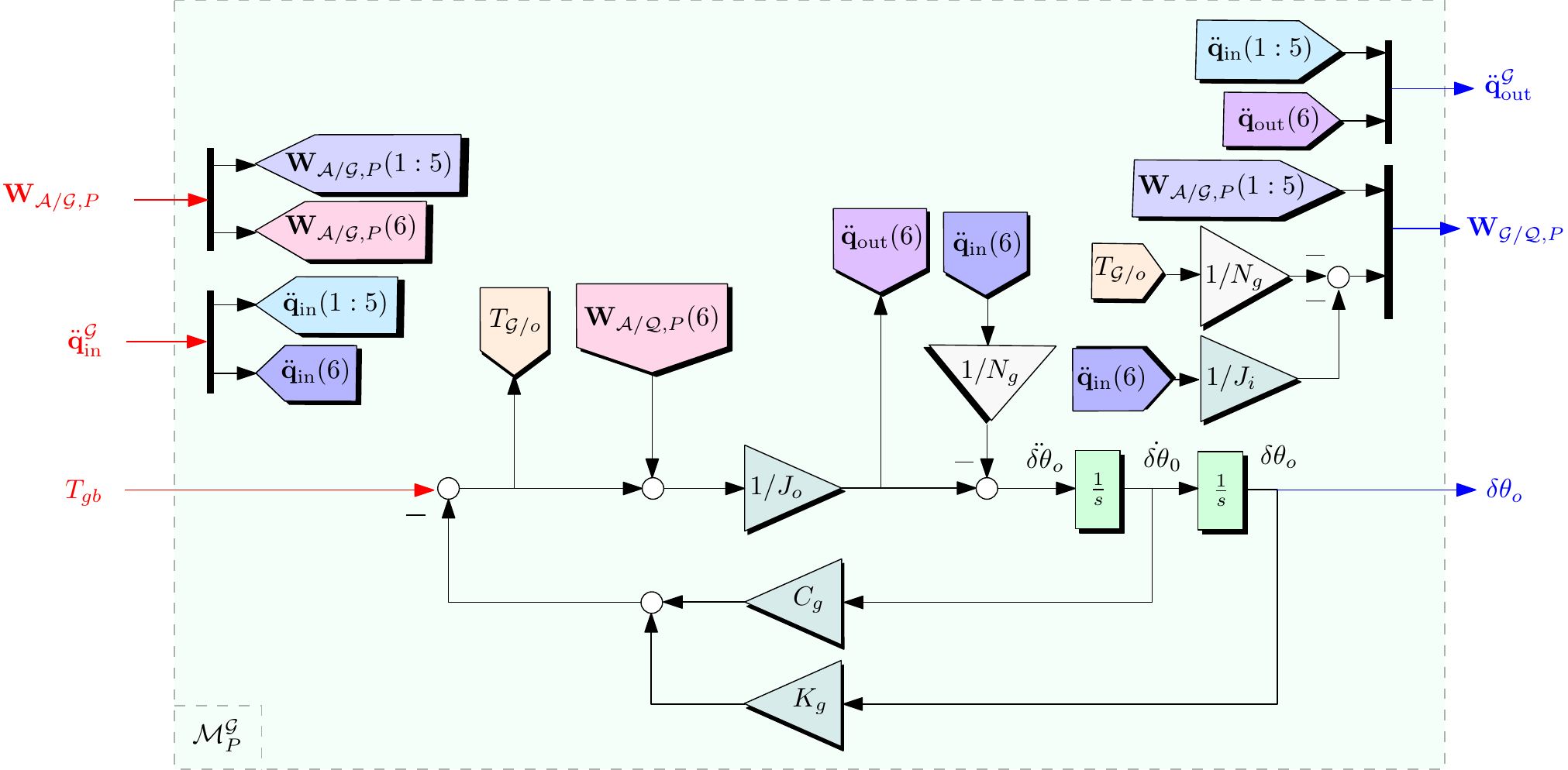} 
	\caption{Gearbox TITOP block diagram with revolute joint axis $\mathbf{z}_g$ aligned with $\mathbf{z}_a$.} 
	\label{fig:gearbox_titop_block_diagram}
\end{figure}

The TITOP model $\mathcal{M}_P^{\mathcal{G}}$ of the gearbox in case $\mathbf{z}_g\equiv\mathbf{z}_a$ (where $z-$axis of the gearbox reference frame $\mathcal{R}_g = (P;\,\mathbf{x}_g,\,\mathbf{y}_g,\,\mathbf{z}_g)$ is the revolute axis) can be described by Eq. \eqref{eq:gb_titop} or equivalently by the block diagram in Fig. \ref{fig:gearbox_titop_block_diagram}. Note that one can chose $J_i = 0$ to implement this model but $J_o \neq 0$.

\section{Model of SADM harmonic disturbance}
\label{sec:SADM_disturbance}
		 The SADM is source of pointing jitter because of three main reasons: the micro-stepping input signal $i$, the detent torque $T_d$ and the defaults of the gearbox gear teeth. 
		 In this section a linear model of the three perturbation sources is provided in order to make possible robust performance analysis as done in Section \ref{sec:wc_pointing_analysis}.  
		 
\subsection{Microstepping input signal}
\label{sec:musignal}
		 The input signal to both systems is a staircase signal of growing integer numbers and stair time-span equal to a $\mu$-step. The period $T_\mu$ of this signal thus takes the expression:
		 \begin{equation}
		 	T_\mu = \frac{2\pi}{\Omega_q z p n_\mu},
		 	\label{eq:T_mu}
		 \end{equation}
	 with $\Omega_q$ the demanded rotor angular speed in rad/s and $n_\mu$ the number of microsteps. Note that if a gearbox is installed between the stepper motor and the appendage $\mathcal{A}$, $\Omega_q$ has to be multiplied by the gear ratio $N_g$ in Eq. \eqref{eq:T_mu}.
	 
\begin{figure}
	 	\centering
	 	\includegraphics[width=\columnwidth]{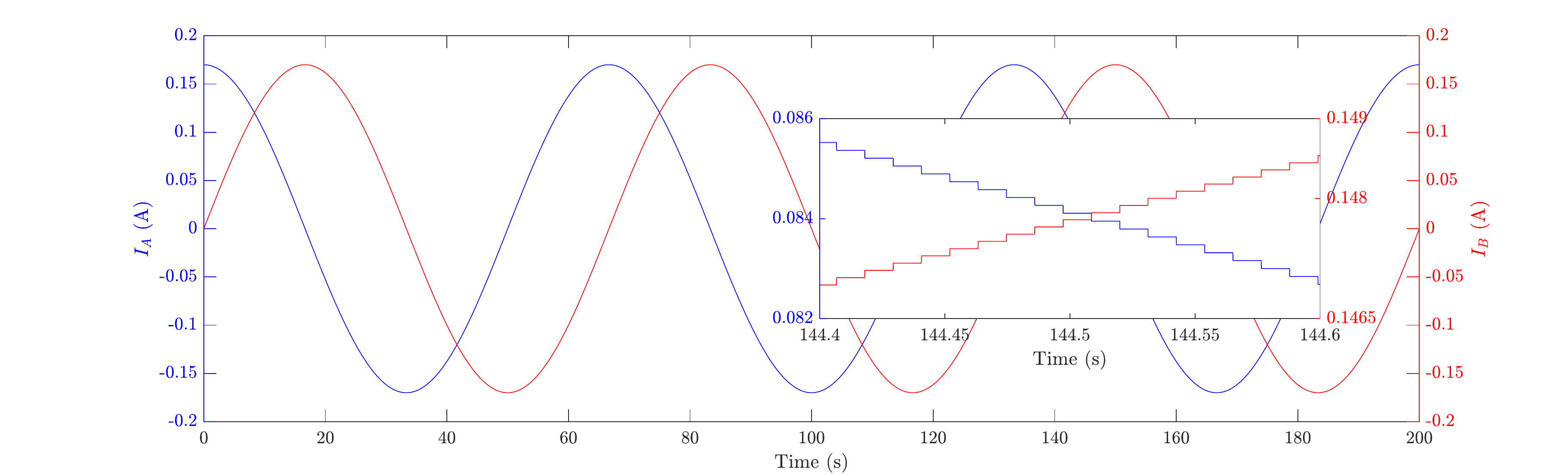} 
	 	\caption{Phase currents}
	 	\label{fig:currents}
	 \end{figure}

	 A detail of the currents in phases A and B  is shown in Fig. \ref{fig:currents}. The currents in the phases have a stair step sinusoidal shape as in Eq. \eqref{eq:IaIb}.

\begin{figure}
  	\centering
  	\includegraphics[width=\columnwidth]{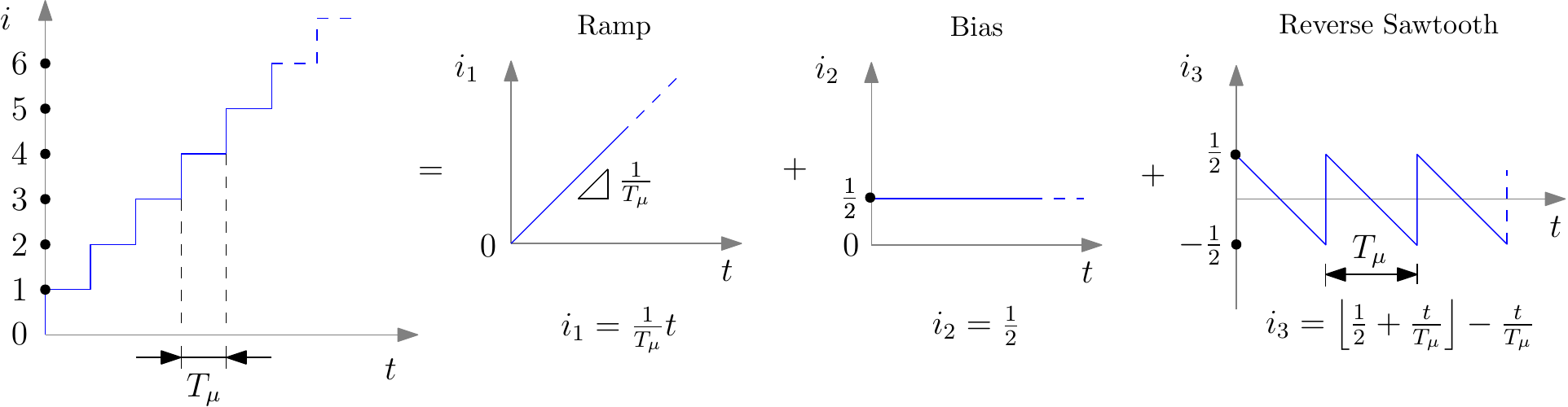} 
  	\caption{Microstep signal generator decomposition}
  	\label{fig:composition_i}
  \end{figure}

  The micro-steps generator signal $i$ can be decomposed in three signals as shown in Fig. \ref{fig:composition_i}: a ramp $i_1$ of slope equal to $1/T_\mu$, a constant bias $i_2 = 1/2$ and a reverse Sawtooth signal $i_3$. Among the three signals, $i_3$ is the source of microvibrations. The other two signals have very low frequency content that is generally damped by the spacecraft AOCS system.
  
  The time expression of $i_3$ is:
  \begin{equation}
  	i_3 = \left\lfloor\frac{1}{2}+\frac{t}{T_\mu}\right\rfloor-\frac{t}{T_\mu},
  	\label{eq:sawtooth}
  \end{equation}
  where $\lfloor\bullet\rfloor$ is the floor operator. This signal can be approximated by a Fourier series as follows:
  \begin{equation}
  	i_3 = \frac{1}{\pi}\sum_{j=1}^{\infty} \frac{1}{h_j}\sin\left(\frac{2\pi h_j}{T_\mu}t\right).
  	\label{eq:sadm_harmonics}
  \end{equation}
  
\begin{figure}[!ht]
	\centering
	\includegraphics[width=\columnwidth]{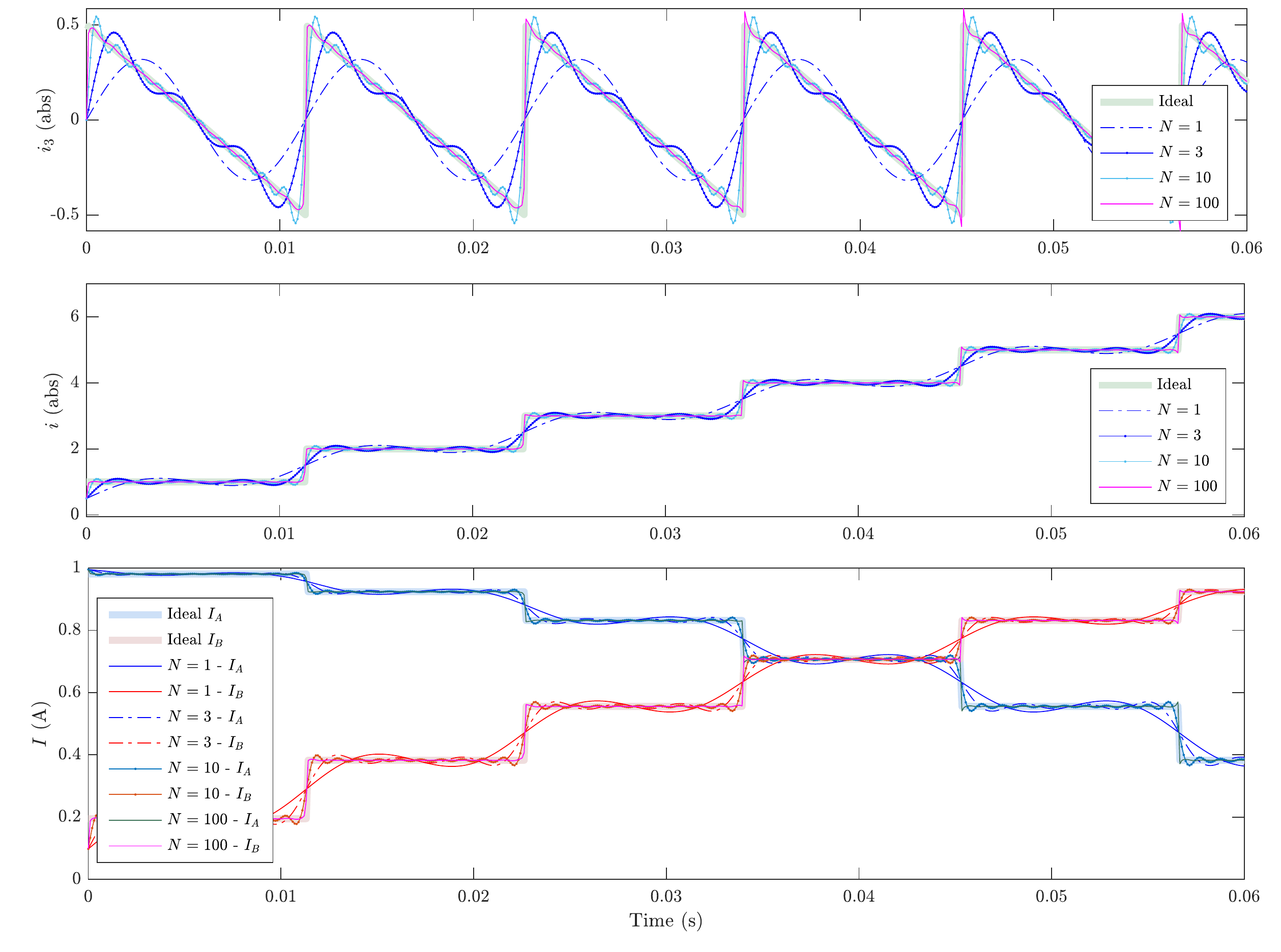} 
	\caption{Fourier approximation with $N$ harmonics: reverse Sawtooth (top), micro-steps generator (middle), phase currents (bottom).} 
	\label{fig:sawtooth_signal}
\end{figure}

If the Fourier series is truncated to the first $N$ terms the resulting reverse Sawtooth signal takes the form illustrated in Fig. \ref{fig:sawtooth_signal} on the top. As shown in Fig. \ref{fig:sawtooth_signal} a small number of harmonics provides a good approximation of the ideal Sawtooth signal in Eq. (\ref{eq:sawtooth}). In the same figure the resulting micro-steps generator $i$ is shown for the same number of harmonics together with the phase currents computed as in Eq. (\ref{eq:IaIb}).
The characterization of the micro-steps generator signal $i$ allows us to reduce the analysis of microvibration induced on the spacecraft to the $N$ disturbing harmonics.

The objective is to find the linear filter:

\begin{equation}
	\mathbf{W}_{\mathrm{hd}} = \left[W_{\mathrm{hd}_1},\,\dots,\,W_{\mathrm{hd}_j},\,\dots,\,W_{\mathrm{hd}_N}\right],
\end{equation}
that outputs the $N$ harmonics in steady-state when driven by the normalized harmonic signals $\mathbf{d}_{i_3}=\left[{d}_{i_{3_1}},\,\dots,\,{d}_{i_{3_j}},\,\dots,\,{d}_{i_{3_N}}\right]^{\mathrm{T}}$, with ${d}_{i_{3_j}} = \sin\left(\frac{2\pi h_j}{T_\mu} t\right)$.

\begin{figure}[!ht]
	\centering
	\includegraphics[width=0.7\columnwidth]{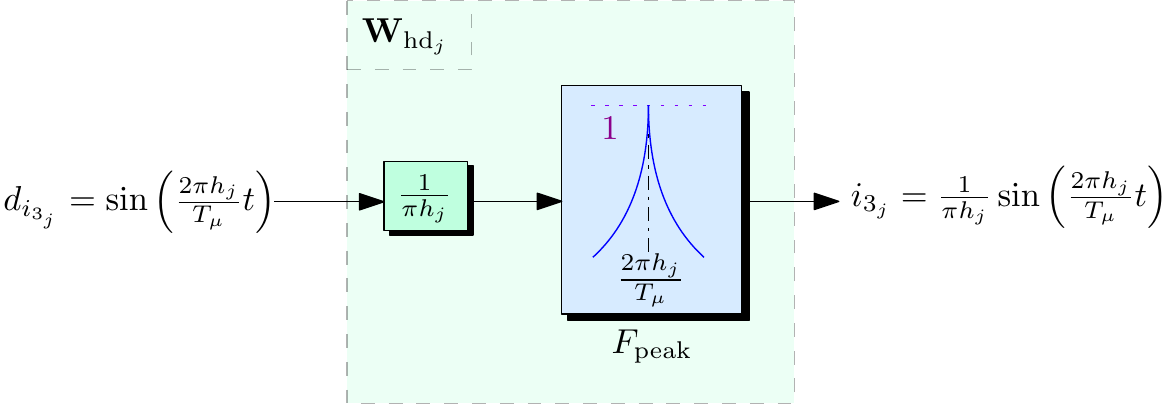} 
	\caption{Structure of the filter ${W}_{\mathrm{hd}_j}$ corresponding to the $j$-th harmonic disturbance.} 
	\label{fig:sadm_weight}
\end{figure}

For each harmonic disturbance the associated filter $W_{\mathrm{hd}_j}$ has the structure illustrated in Fig. \ref{fig:sadm_weight}, where $F_{\mathrm{peak}}$ approximates an harmonic signal of unitary spectral amplitude. $F_{\mathrm{peak}}$ is chosen as a second order unit gain band-pass filter centered at the frequency $2\pi h_j/T_\mu$:
\begin{equation}
	F_{\mathrm{peak}} = \frac{\beta s}{s^2+\beta s + \left(\frac{2\pi h_j}{T_\mu}\right)^2} = \frac{\beta s}{s^2 + \beta s + \left( h_j \Omega_q z p n_\mu\right)^2}, \quad \mathrm{for\, all}\, \beta>0,\,\beta \in \mathbb{R}
	\label{eq:f_peak_sadm}
\end{equation}

where $\beta$ is a parameter that controls the filter bandwidth.

In order to match the amplitude of the harmonic disturbances in Eq. (\ref{eq:sadm_harmonics}), $F_\mathrm{peak}$ has to be multiplied by the factor $1/\left(\pi h_j\right)$ to obtain the expression of the filter $W_{\mathrm{hd}_j}$:

		\begin{equation}
			W_{\mathrm{hd}_j} = \frac{1}{\pi h_j}F_{\mathrm{peak}} 
		\end{equation}

\begin{figure}[!ht]
	\centering
	\includegraphics[width=0.9\columnwidth]{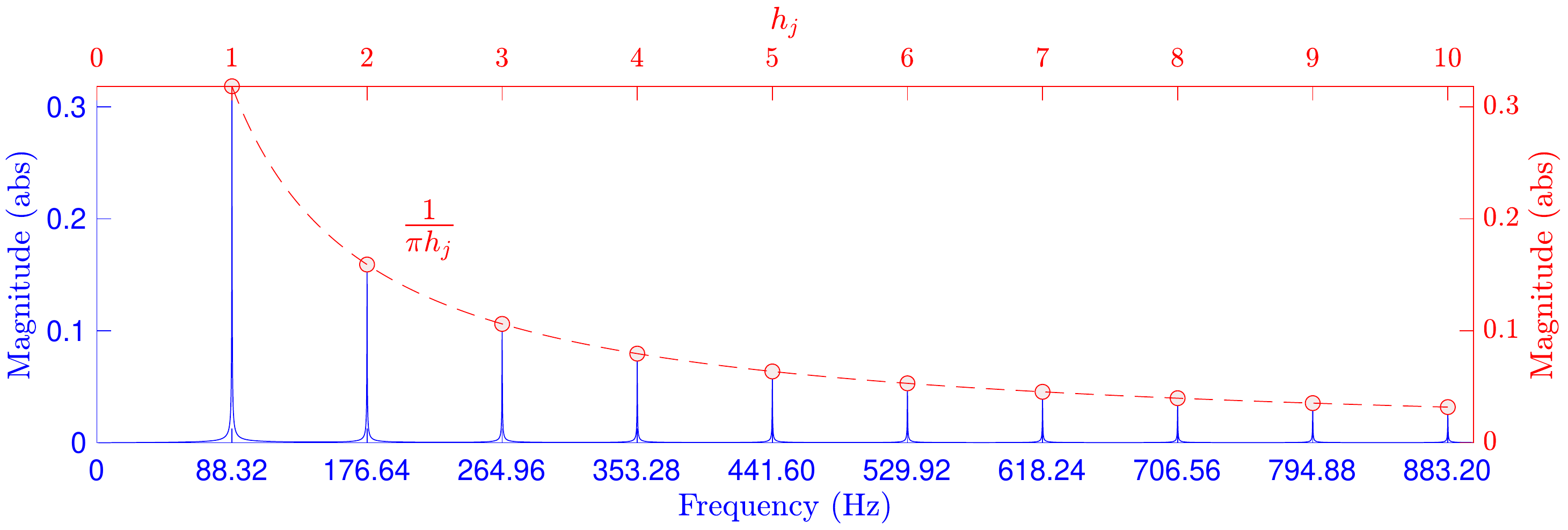} 
	\caption{Magnitude of the weight $\mathbf{W}_{\mathrm{hd}}$ by considering the first 10 harmonics.} 
	\label{fig:sadm_weights}
\end{figure}

For a micro-step generator whose data are detailed in \ref{tab:data} the filter $\mathbf{W}_{\mathrm{hd}}$ representing the first ten harmonics is shown in Fig. \ref{fig:sadm_weights}. Notice as the peak amplitude of each harmonic belongs to the curve $\frac{1}{\pi h_j}$: the amplitude spectrum of the harmonic disturbances multiple of the main harmonic is divided by the corresponding harmonic number. This conclusion can help to choose how many harmonics should be considered for the analysis of the pointing performance without augmenting the number of the states of the entire system. The less important harmonics can then be condensed in a broadband noise.

\subsection{Detent torque}
\label{sec:model_detent_torque}

As already stated in Section \ref{sec:step_motor} the detent torque $T_d$ is generally considerably small compared to the motor torque $T_m$. However an exhaustive vibration analysis requires to deal with this disturbing source of harmonic nature. The term relative to $T_d$ in Eq. \eqref{eq:te_motor} cannot be linearized but it can be considered as an external sinusoidal disturbance that acts directly on the SADM rotor. The detent torque presents a fundamental periodicity $T_{4\theta_r}$ that is a quarter of the fundamental period of the motor torque as stated by Eq. \eqref{eq:te_motor}:
\begin{equation}
	T_{4\theta_r} = \frac{2\pi}{4\Omega_q z}.
	\label{eq:T_det}
\end{equation}

Note that if a gearbox is installed between the stepper motor and the appendage $\mathcal{A}$, $\Omega_q$ has to be multiplied by the gear ratio $N_g$ in Eq. \eqref{eq:T_det}.

The periodicity $T_{\theta_r}$ of the mechanical angle $\theta_r$ is in fact $n_\mu p$ times the periodicity $T_\mu$ of the electrical $\mu$-step electrical angle.

As stated in \cite{sanfe2019}, only the first harmonic of this signal can be analytically determined. Its amplitude is in fact equal to the detent torque constant $K_d$. For the successive harmonics multiple of the fundamental one an experimental identification is necessary. In this work we consider that the entire perturbation coming from the detent torque $T_d$ can be approximated as a signal similar to $i_3$:
\begin{equation}
	T_d = K_d\sum_{j=1}^{\infty}\frac{1}{h_j}\sin\left(\frac{2\pi h_j}{T_{4\theta_r}}t\right)
\end{equation} 

We follow the same reasoning of Section \ref{sec:musignal} to find the linear filter $\mathbf{W}_{d_{hd}} $ that models $N$ disturbing harmonics:

\begin{equation}
	\mathbf{W}_{d_{hd}} = \left[W_{\mathrm{d_{hd}}_1},\,\dots,\,W_{\mathrm{d_{hd}}_j},\,\dots,\,W_{\mathrm{d_{hd}}_N}\right],
\end{equation}

that outputs the $N$ harmonics in steady-state when driven by the normalized harmonic signals
\begin{equation*}
	\mathbf{d}_{T_d}=\left[{d}_{T_{d_1}},\,\dots,\,{d}_{T_{d_j}},\,\dots,\,{d}_{T_{d_N}}\right]^{\mathrm{T}}\quad \mathrm{with}\quad {d}_{T_{d_j}} = \sin\left(\frac{2\pi h_j}{T_{4\theta_r}} t\right).
\end{equation*}

For each harmonic disturbance the associated filter $W_{\mathrm{d_{hd}}_j}$ has the structure:
\begin{equation}
		W_{\mathrm{d_{hd_j}}} = \frac{K_d}{h_j} \frac{\beta s}{s^2 + \beta s + \left( 4h_j \Omega_q z\right)^2}
\end{equation}
		 
\subsection{Modeling imperfections in the gearbox teeth}
\label{sec:model_gearbox_damage}

\begin{wrapfigure}[18]{r}{0.5\columnwidth}
	\begin{center}
		\vspace{-20pt}
		\includegraphics[width=.5\columnwidth]{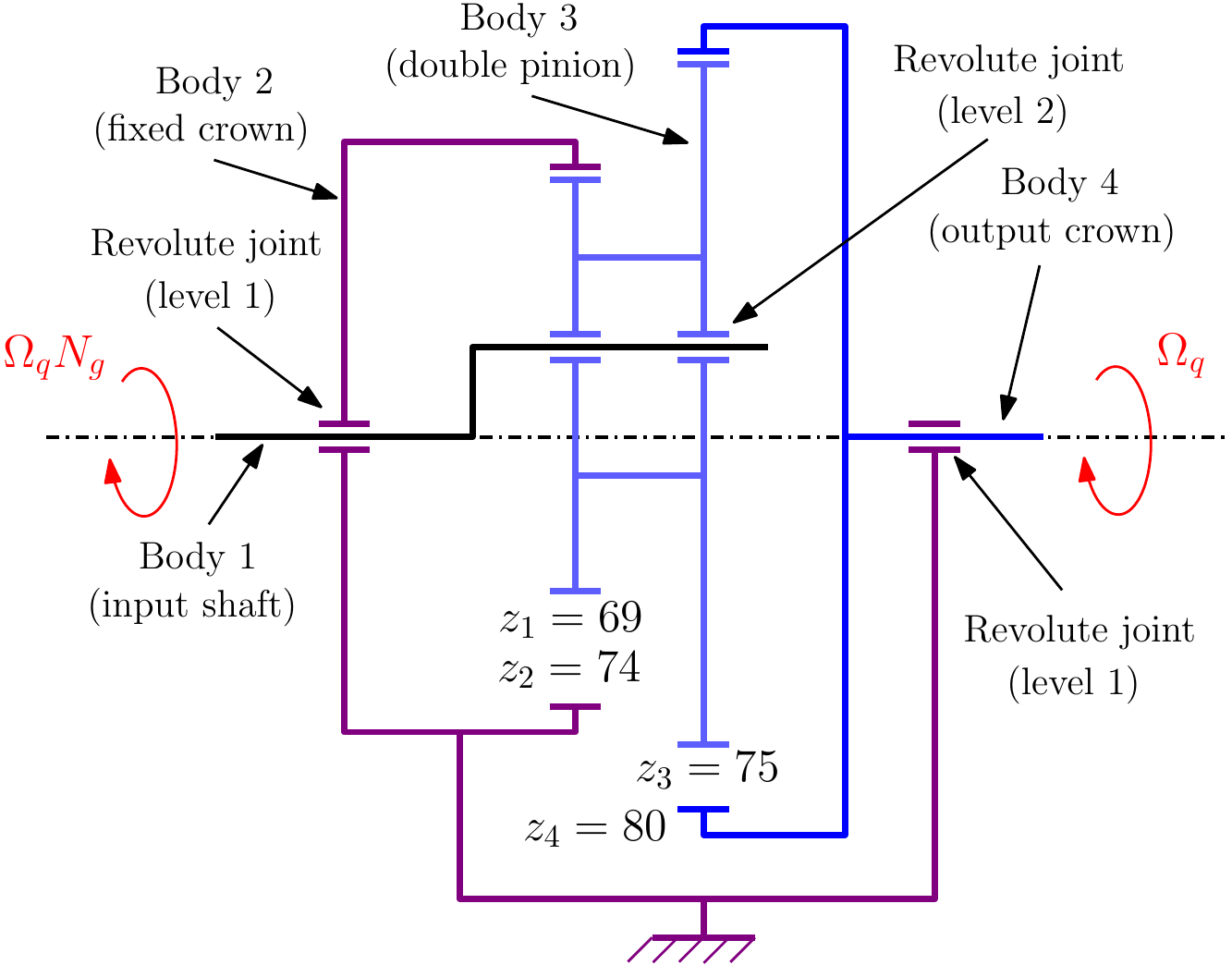}
	\end{center}
	\vspace{-10pt}
	\caption{Kinematic model of the SADM gearbox ($z_j$: gear teeth number)}
	\label{fig:gearbox_scheme}
\end{wrapfigure}

From one side the use of a gearbox shifts the fundamental harmonics both of the micro-stepping and detent torque signals to higher frequencies using the gear ratio $N_g$ that divides both the periods in Eqs. \eqref{eq:T_mu}	and \eqref{eq:T_det}. On the other side the gearbox introduces supplementary harmonics due to the frequency of the contact imperfections in its gear pairs at lower frequencies.
This is critical if natural modes of the solar panel are present in the same frequency bandwidth.
A toolbox previously developed by the authors \cite{Penaud2012,al2021} allows to compute the frequencies $\Omega_{d_j}$ (rad/s) of the $j-$th contact imperfection in any gear pairs of any mechanism. The toolbox requires only the description of the mechanism as it is depicted in Fig. \ref{fig:gearbox_scheme}.

Then the toolbox can compute the angular rate $\Omega_j$ of each $j-$th body involved in the mechanism (see Table \ref{tab:gear_omega}) and the characterization of each contact imperfection in the various gear pairs (see Table \ref{tab:gear_damage_freq}: 6 imperfections can be identified in the SADM gearbox of Fig. \ref{fig:gearbox_scheme}). Note that $\Omega_{d}^j$ is the frequency in rad/s of the $j-$th gearbox imperfection.

\begin{table}[h!]
	\centering
	\begin{tabular}{ccccc}
		\hline
		\textbf{Body} & 1 & 2 & 3 & 4 \\ \hline
		\textbf{$\Omega_j/\Omega_q$} & -184 & 0 & 13.33 & 1 \\ \hline
	\end{tabular}
\caption{Body angular rates in the gearbox mechanism normalized according to the output shaft (body 4)}
\label{tab:gear_omega}
\end{table}

\begin{table}[h!]
	\centering
       {\begin{tabular}{ccccc}
		\hline
		\textbf{Imperfection} & $\bm{\tilde{\Omega}_{d}^j={\Omega}_{d}^j/\Omega_q}$ & \textbf{Gear Pair} & \textbf{Cause} & \textbf{Number of sources} \\ \hline
		1 & 13616 & (2,3) & gear frequency & 1 \\
		2 & 184 & (2,3) & a tooth on body 2 & 74 \\
		3 & 197.3 & (2,3) & a tooth on body 3 & 69 \\
		4 & 14800 & (3,4) & gear frequency & 1 \\
		5 & 197.3 & (3,4) & a tooth on body 3 & 75 \\
		6 & 185 & (3,4) & a tooth on body 4 & 80 \\ \hline
	\end{tabular}}
	\caption{Contact imperfection frequency in the gearbox mechanism, normalized according to the output shaft (body 4)}
	\label{tab:gear_damage_freq}
\end{table}

Each $j-$th imperfection  can create an harmonic disturbing torque $T_{gb}^j$ that has a Sawtooth profile involving the fundamental frequency $\Omega_{d}^j$. The fundamental periodicity $T_{dh}^j$:
\begin{equation}
	T_{dh}^j = \frac{2\pi}{\Omega_{d}^j}.
	\label{eq:T_dh}
\end{equation}

The torque $T_{gb}^j$ takes thus the form:

\begin{equation}
	T_{gb}^j = \frac{1}{\pi}\sum_{k=1}^{\infty} \frac{1}{h_k}\sin\left( \frac{2\pi h_k}{T_{dh}^j} t\right).
	\label{eq:gearbox_sawtooth}
\end{equation}

By following the same raisoning of Section \ref{sec:musignal} to find the linear filter $\mathbf{W}_{gb}^j $ that models $N$ disturbing harmonics of the $j-$th imperfection:

\begin{equation}
	\mathbf{W}_{gb}^j = \left[W_{gb_1}^j,\,\dots,\,W_{gb_k}^j,\,\dots,\,W_{gb_N}^j\right],
\end{equation}

that outputs the $N$ harmonics in steady-state when driven by the normalized harmonic signals
\begin{equation*}
	\mathbf{d}_{T_{gb}}^j=\left[{d}_{T_{gb_1}}^j,\,\dots,\,{d}_{T_{gb_k}}^j,\,\dots,\,{d}_{T_{gb_N}}^j\right]^{\mathrm{T}}\quad \mathrm{with}\quad {d}_{T_{gb_k}}^j = \sin\left(\frac{2\pi h_k}{T_{dh}^j} t\right).
	\label{eq:T_{dh}}
\end{equation*}

For each $k-$th harmonic disturbance of the $j-$th gearbox imperfection the associated filter $W_{gb_k}^j$ has the structure:
\begin{equation}
	W_{gb_k}^j = \frac{1}{\pi h_k} \frac{\beta s}{s^2 + \beta s + \left( h_j \Omega_d^j \right)^2}
\end{equation}

\begin{figure}[th!]
	\centering
	\includegraphics[width=.8\columnwidth]{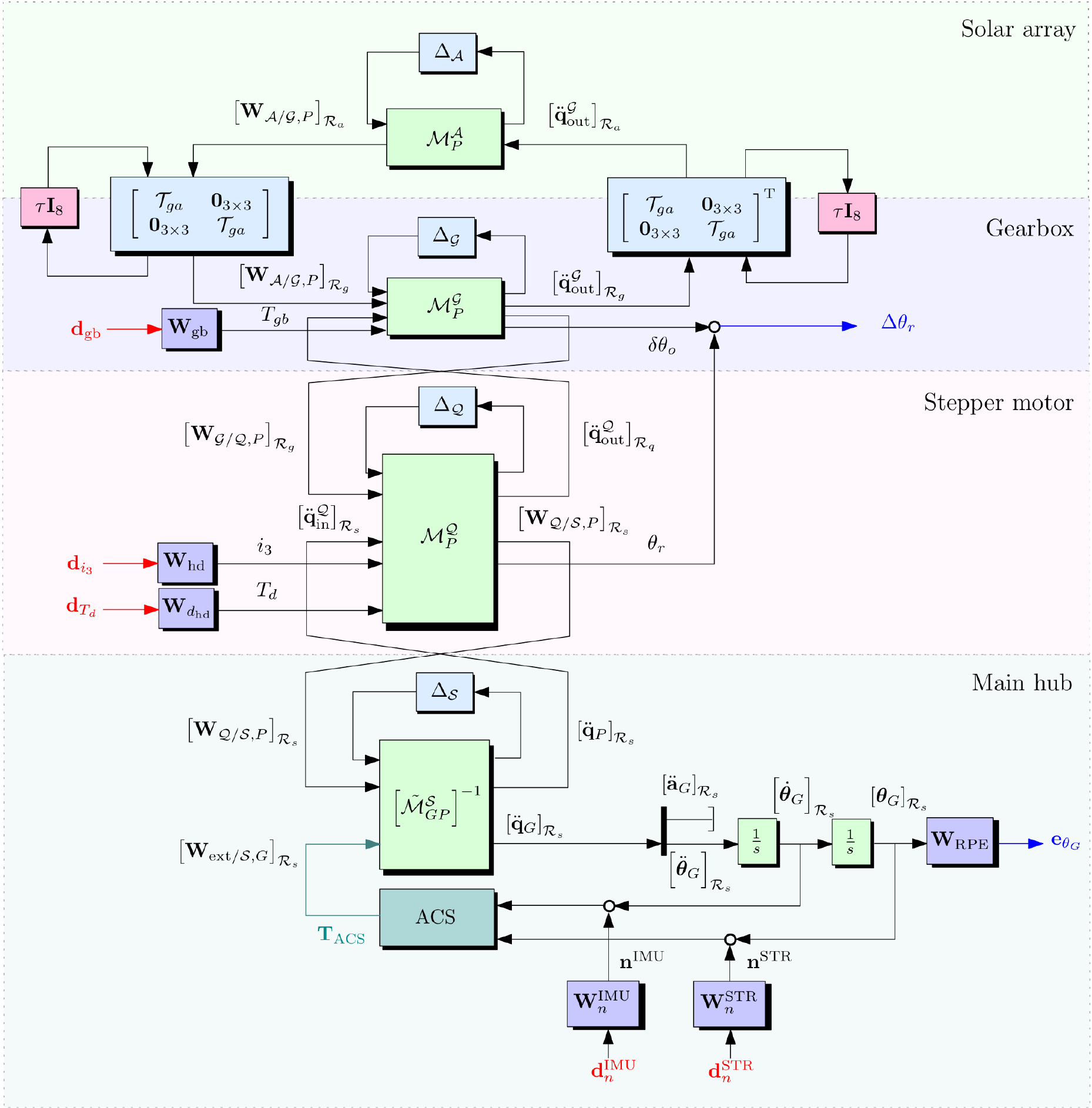} 
	\caption{Connection diagram of a flexible spacecraft with SADM}
	\label{fig:lpv_sadm_satellite}
\end{figure}
		 
\section{Spacecraft model validation}
\label{sec:model_validation}
In this section a European spacecraft is used as study case to build the entire system model. The mechanical properties data are summarized in \ref{tab:data}. 

By connecting all the blocks modeling a central body (Section \ref{sec:model_central_body}), a flexible solar array (Section \ref{sec:model_appendage}), a stepper motor (Section \ref{sec:step_motor}) and a reduction gearbox (Section \ref{sec:model_gearbox}) together with the three perturbation sources introduced in Section \ref{sec:SADM_disturbance} and tacking into account the rotation of the solar array (Section \ref{sec:connection}), it is possible to obtain the connection block diagram in Fig. \ref{fig:lpv_sadm_satellite}. 

 Note that the unit variance Gaussian noise of the IMU and the star tracker are respectively scaled with the weighting filters $\mathbf{W}_n^\mathrm{IMU}$ and $\mathbf{W}_n^\mathrm{STR}$.
The ACS is based on a simple decoupled  proportional-derivative controller tuned on the total static inertia of the spacecraft as proposed in \cite{Guy2014}.  
If $I_{xx}^{\mathcal{S}+\mathcal{Q}+\mathcal{G}+\mathcal{A}}$ is the static inertia of the entire spacecraft on x-axis, the proportional and derivative gains (respectively $K_{p_x}$ and $K_{v_x}$) are computed as it follows:
\begin{equation}
	K_{p_x} = \omega_{\mathrm{ACS}}^2 I_{xx}^{\mathcal{S}+\mathcal{Q}+\mathcal{G}+\mathcal{A}}, \quad K_{v_x} = 2 \zeta_{\mathrm{ACS}} \omega_{\mathrm{ACS}} I_{xx}^{\mathcal{S}+\mathcal{Q}+\mathcal{G}+\mathcal{A}},
\end{equation}
where $\omega_{\mathrm{ACS}}$ and $\zeta_{\mathrm{ACS}}$ are respectively the demanded closed-loop bandwidth and damping ratio.
The same approach is used for the axis $y$ and $z$.
The static control law results then $\mathbf{T}_{\mathrm{ACS}} = - \mathbf{K}_{\mathrm{ACS}} \left[ \dot{\bm{\theta}}_G\,{\bm{\theta}}_G\right]$, where:
\begin{equation}
	\mathbf{K}_{\mathrm{ACS}} =  \left[ \mathrm{diag}(K_{v_x},K_{v_y},K_{v_z}),\mathrm{diag}(K_{p_x},K_{p_y},K_{p_z})\right]
\end{equation} 

The ACS design parameters chosen for the study case are $\omega_{\mathrm{ACS}} = 0.01\,\mathrm{rad/s}$ and $\zeta_{\mathrm{ACS}} = 0.7$ for the three axes.

The attitude performance is expressed in terms of Relative Pointing Error (RPE) as defined in \cite{Ott}. The RPE performance corresponds to the angular difference between the instantaneous pointing vector $[\bm{\theta}_G]_{\mathcal{R}_s}$ and the short-time average pointing vector during a given time period $t_\Delta$.
In frequency domain the RPE performance corresponds to the high-pass performance weight \cite{Ott2013}:
\begin{equation}
	\mathbf{W}_{\mathrm{RPE}} = \epsilon_{max}^{-1}\frac{t_\Delta s \left(t_\Delta s+\sqrt{12}\right)}{\left(t_\Delta s\right)^2+6\left(t_\Delta s\right)+12}\mathbf{I}_3,
\end{equation}
where $\epsilon_{max}$ overbounds the maximum RPE target value.
For the present study case $t_\Delta = 3\,\mathrm{ms}$ and $\epsilon_{max} = 0.06 \,\mathrm{\mu rad/\sqrt{Hz}}$. 

\begin{wrapfigure}[25]{r}{0.5\columnwidth}
	\begin{center}
		\vspace{-20pt}
		\includegraphics[width=.5\columnwidth]{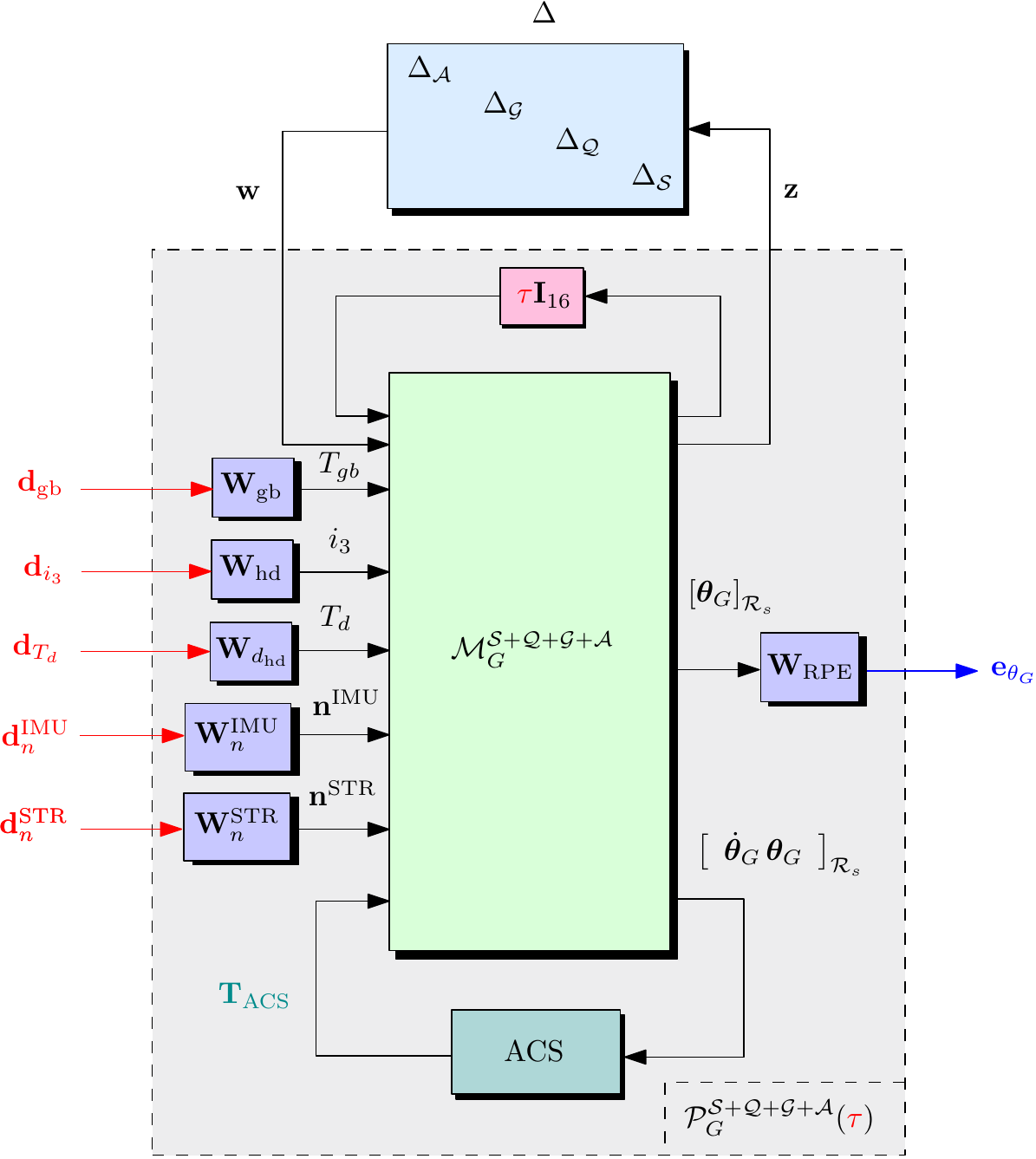}
	\end{center}
	\vspace{-10pt}
	\caption{LFT diagram of the spaceraft system}
	\label{fig:LPV_LFT}
\end{wrapfigure}

The assembling of the entire system as showed in Fig. \ref{fig:lpv_sadm_satellite} can be done in \textsc{Simulink} environment thanks to the new release of the SDT \cite{alaSanf2021}.
The advantage to use the SDT is that the parameters in each sub-block can be declared as uncertain. The system is then already set in minimal LFT form and it can be directly used for linear robust controller synthesis and robust stability/performance assessment. All the features of the \textsc{Matlab} \textit{Robust Control Toolbox} are then directly applicable to the system.
The system in Fig. \ref{fig:lpv_sadm_satellite} can then easily be put in the LFT generalized form in Fig. \ref{fig:LPV_LFT}. The resulting system $\mathcal{F}_u\left(\mathcal{P}_{G}^{\mathcal{S}+\mathcal{Q}+\mathcal{G}+\mathcal{A}}(\textcolor{red}{\tau}),\Delta \right)$ is an uncertain LPV system dependent on the parameter $\tau$ that varies with the rotor angular configuration. $\mathcal{P}_{G}^{\mathcal{S}+\mathcal{Q}+\mathcal{G}+\mathcal{A}}(\textcolor{red}{\tau})$ represents the closed-loop generalized LPV plant between all the input disturbance channels and the performance output channel represented by the RPE of the spacecraft attitude.

Let analyse the nominal transfer function (no uncertainties are taken into account) $i_3 \rightarrow \bm{\theta}_G$ from the stepper motor harmonic disturbance $i_3 $ to the attitude vector $\bm{\theta}_G$ as function of the rotor angular configuration $\theta_r$.
The first observation is that the five solar array modes are concentrated in the frequency bandwidth $\left[0.3 \div 5\right]\,\mathrm{Hz}$. Moreover we note that the first and second mode vary in amplitude and frequency according to $\theta_r$.
The gearbox introduces a flexible mode at $6.052\,\mathrm{Hz}$ and the stepper motor another one at high frequency ($125.4\,\mathrm{Hz}$). 
Note that all these natural frequencies can also vary if the mechanical parameters are uncertain as it is in the reality. We treat the uncertain plant in Section \ref{sec:wc_pointing_analysis}.

\begin{figure}[h!]
	\centering
	\includegraphics[width=.9\columnwidth]{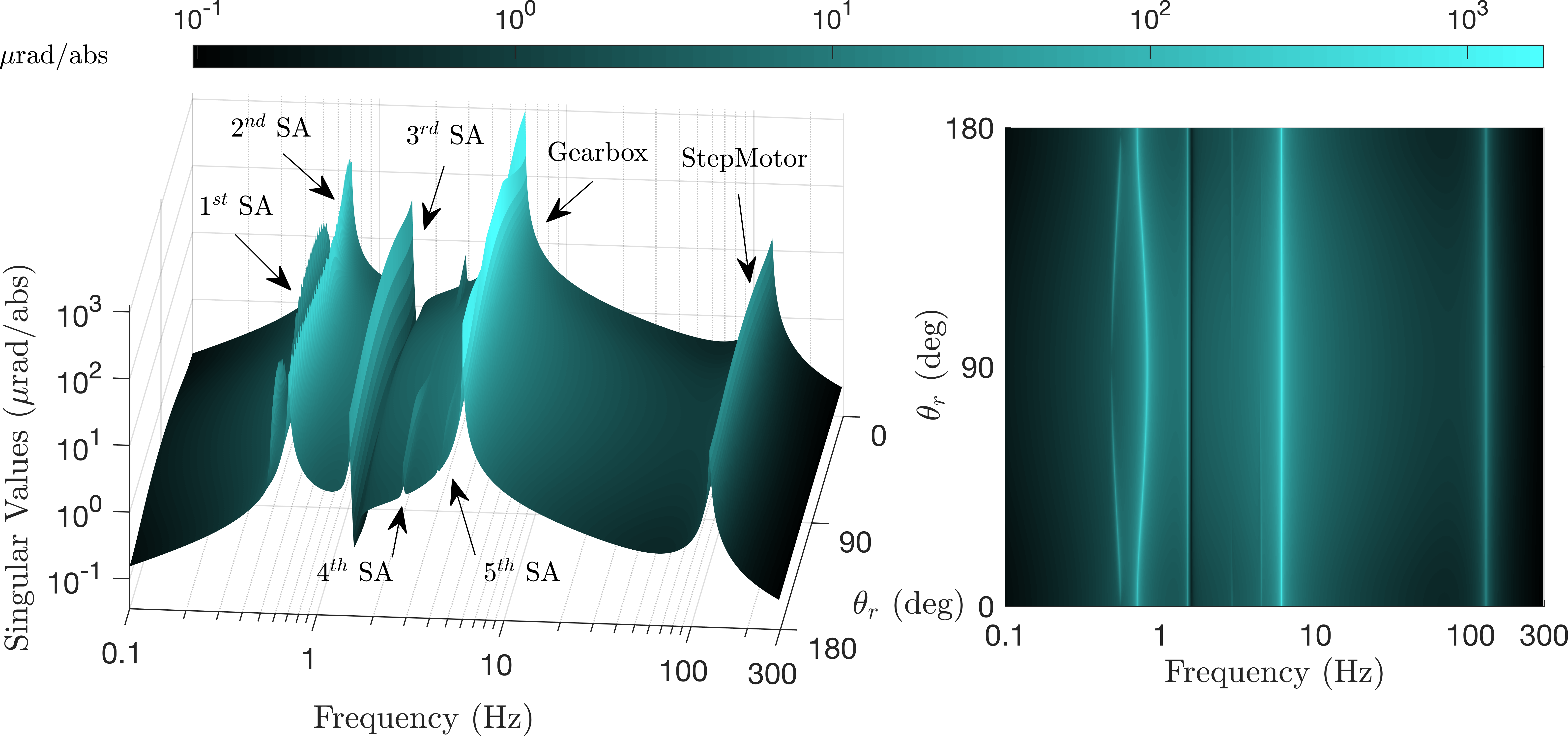} 
	\caption{Singular values of the nominal transfer function $i_3\rightarrow \bm{\theta}_G$ for $\theta_r\in\left[0,180\right]^{\circ}$}
	\label{fig:transf_ustep_theta}
\end{figure}

The problem raises if some SADM harmonics excite the system at these particular frequencies. 
By using Eqs. \eqref{eq:T_mu}, \eqref{eq:T_det} and \eqref{eq:T_dh} it is possible to compute the fundamental frequency of each of the presented harmonic disturbances coming from SADM. Note that due to the presence of the gearbox, the value of the demanded rotor speed $\Omega_q$ needs to be multiplied by the gearbox ratio $N_g$.
Table \ref{tab:fundamental_freq} outlines the results of this computation and considers the 2 typical values
of the Solar Array angular rates $\Omega_q$: the nominal Sun pointing mode (slow) and the rewind mode (fast)..

\begin{table}[h!]
	\centering
	\begin{tabular}{ccc}
		\hline
		\textbf{Fundamental Frequency} & $\bm{\Omega_q = 0.06\,\mathrm{^\circ/s}}$ & $\bm{\Omega_q = -0.205\,\mathrm{^\circ/s}}$ \\ \hline
		{Microstepping} & \pgfsetfillopacity{0.2}\colorbox{YellowGreen}{\pgfsetfillopacity{1}88.320 Hz} & \pgfsetfillopacity{0.2}\colorbox{YellowGreen}{\pgfsetfillopacity{1}301.76 Hz} \\
		{Detent Torque} & \pgfsetfillopacity{0.2}\colorbox{YellowGreen}{\pgfsetfillopacity{1}11.040 Hz}& \pgfsetfillopacity{0.2}\colorbox{YellowGreen}{\pgfsetfillopacity{1}37.720 Hz} \\
		{Gearbox imperfection} $\tilde{\Omega}_d = 184$ & \pgfsetfillopacity{0.2}\colorbox{Melon}{\pgfsetfillopacity{1}0.0307 Hz} & \pgfsetfillopacity{0.2}\colorbox{Melon}{\pgfsetfillopacity{1}0.1048 Hz}\\
		{Gearbox imperfection} $\tilde{\Omega}_d = 197.3$ & \pgfsetfillopacity{0.2}\colorbox{Melon}{\pgfsetfillopacity{1}0.0329 Hz} & \pgfsetfillopacity{0.2}\colorbox{Melon}{\pgfsetfillopacity{1}0.1124 Hz} \\ \hline
	\end{tabular}
	\caption{Fundamental frequencies of the SADM disturbances}
	\label{tab:fundamental_freq}
\end{table}

The gearbox imperfections are the main contributors to the excitation of the flexible structures by representing the main concern for pointing performance. Micro-stepping and detent torque act at higher frequencies where the transfer function $\left[\mathbf{d}_{i_3}^\mathrm{T}\,\mathbf{T}_{d}^\mathrm{T} \right]^\mathrm{T}\rightarrow \mathbf{e}_{\bm{\theta}_G}$ rolls off to low gain. Moreover the small magnitude of the detent torque (assuming $K_d\approx 0.2 K_m I$) with respect to the gearbox imperfections makes this transfer negligible to the overall pointing budget.  
 
\begin{figure}[h!]
	\centering
	\includegraphics[width=\columnwidth]{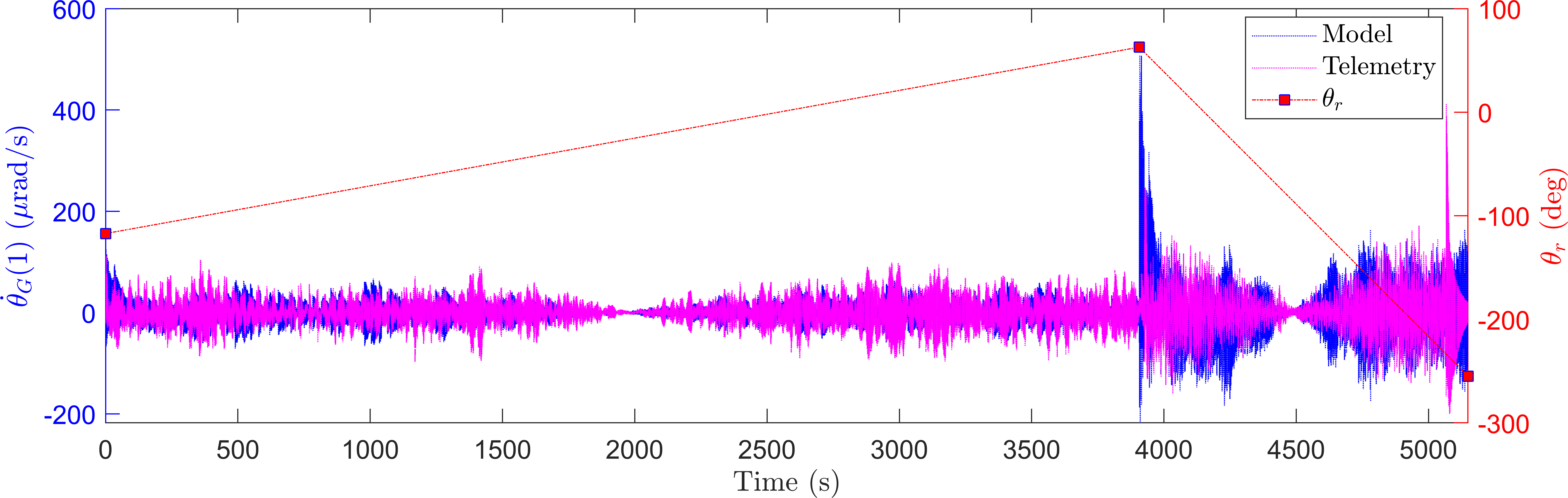} 
	\caption{Validation of the model: comparison with telemetry data on the spacecraft angular speed about $x$-axis with the corresponding evolution of the rotor angular configuration $\theta_r$}
	\label{fig:comparison_telemetry}
\end{figure}

Thanks to this analysis it is now possible to simulate the system by injecting 90 harmonics of the two gearbox imperfections numbered 2 and 3 in Table \ref{tab:gear_damage_freq}. The analysis is focused on these two imperfections due to the possible matching of their harmonic frequencies with the flexible mode frequencies. The simulation results are then compared with the spacecraft on-orbit available IMU telemetries. The system is simulated at $10\,\mathrm{Hz}$ for a period of $5150\,\mathrm{s}$ (one orbit). At the beginning the solar array starts from an angular position $\theta_r(0) = -117.2\,\mathrm{^\circ}$ and it is commanded to rotate at slow rate of $\Omega_q = 0.06\,\mathrm{^\circ/s}$ for $2\theta_r(0)/\Omega_q$ seconds. Then a fast rotor rate of $\Omega_q = -0.205\,\mathrm{^\circ/s}$ is maintained till the end of the simulation.
The comparison of the simulation with the on-orbit telemetries on $x-$axis is shown in Fig. \ref{fig:comparison_telemetry}. We see that in time domain the LPV nominal model manages to capture the evolution of the rate pointing error $\dot{\theta}_G(1)$ with the changing of the solar array angular position $\theta_r$.

\begin{figure}[h!]
	\centering
	\includegraphics[width=\columnwidth]{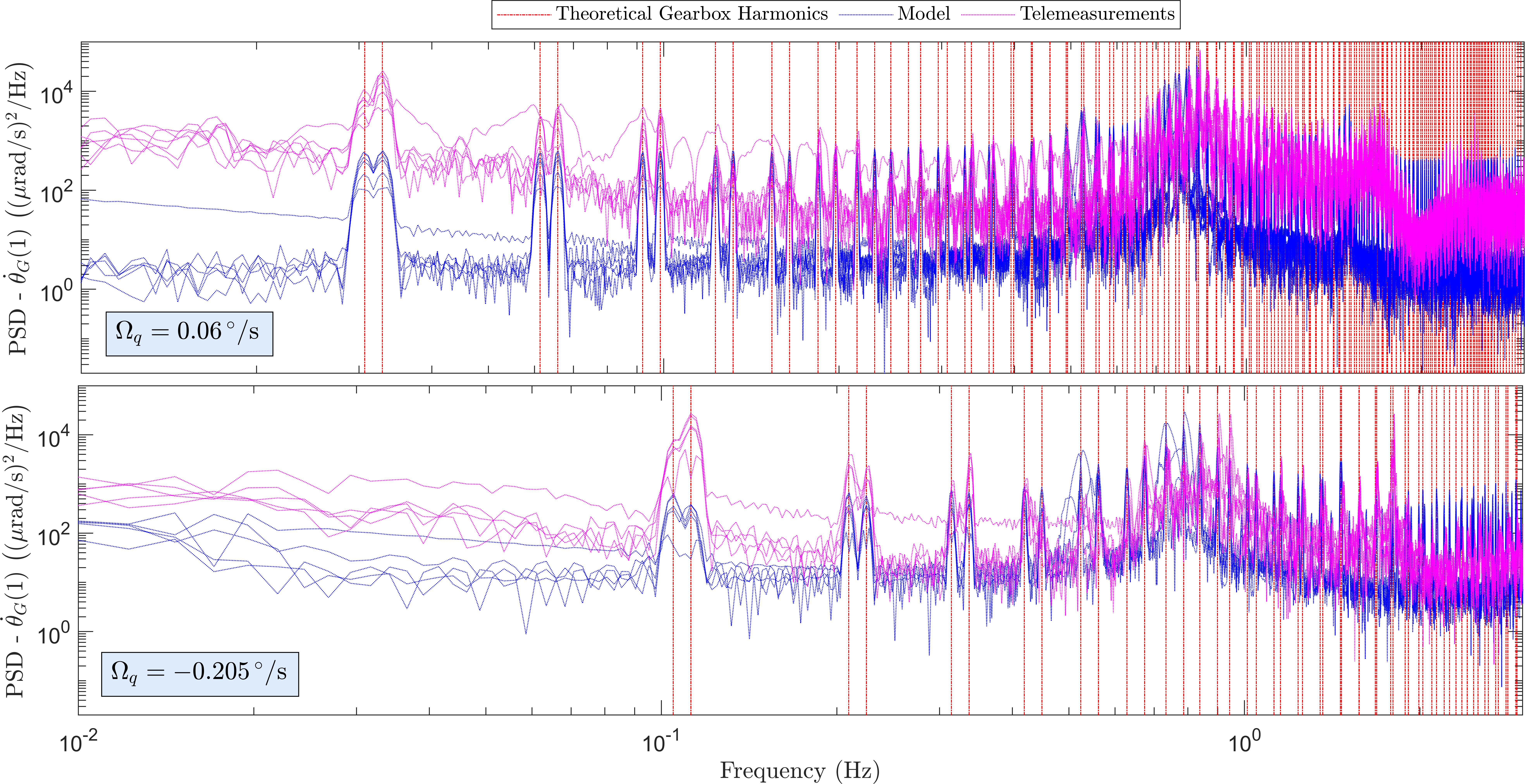} 
	\caption{Validation of the model: comparison with the spectrogram of telemetry data on the spacecraft angular speed about $x$-axis for slow rotor rate (top) and fast rotor rate (bottom)}
	\label{fig:PSD_slow_fast}
\end{figure}

However what is more interesting is the comparison in frequency domain in Fig. \ref{fig:PSD_slow_fast}. A Spectrogram is applied to two different set of data samples of both the simulation and telemetry data in order to capture the evolution of the excited frequencies for both slow rotor rate and fast rotor rate.
The two sets of data are reduced to the samples where the behavior is stationary (due to the settling time of the ACS) in order to avoid the big peaks in Fig. \ref{fig:comparison_telemetry} due to the transitory.
We notice that the model perfectly captures all the interesting frequencies stored in telemetry. The amplitude difference of the peaks at lower frequencies is mainly due to a miss-knowledge of the authors of the actual control law implemented on the real spacecraft. 

\section{Worst-case pointing analysis}\label{sec:wc_pointing_analysis}

Once the model is validated by on-orbit telemetry it is possible to make deeper worst-case analysis for the system in Fig. \ref{fig:LPV_LFT} to assess the robustness of the required pointing performance. This is done by computing the bounds of the worst case $\mathcal{H}_\infty$-norm for the transfer function $\left[\mathbf{d}_{i_3}^\mathrm{T}\,\mathbf{T}_{d}^\mathrm{T} \right]^\mathrm{T}\rightarrow \mathbf{e}_{\bm{\theta}_G}(\bullet)$ across $\tau \in \bm{\tau}$, where $\bm{\tau}$ is a vector of $N_\tau$ discrete values of the parameter $\tau$. This corresponds to compute the structured singular value upper bound
\begin{equation}
	\mathrm{sup}_\Delta\mu_\Delta = \mathrm{sup}_\Delta\left\vert\left\vert\mathcal{F}_u\left( \mathcal{P}_G^{\mathcal{S}+\mathcal{Q}+\mathcal{G}+\mathcal{A}}(\textcolor{red}{\tau}),\Delta\right)_{tf}\right\vert\right\vert_\infty,
\end{equation}
and the lower bound
\begin{equation}
	\mathrm{inf}_\Delta\mu_\Delta = \mathrm{inf}_\Delta\left\vert\left\vert\mathcal{F}_u\left( \mathcal{P}_G^{\mathcal{S}+\mathcal{Q}+\mathcal{G}+\mathcal{A}}(\textcolor{red}{\tau}),\Delta\right)_{tf}\right\vert\right\vert_\infty,
\end{equation}
for the channel $tf:\,\left[\mathbf{d}_{i_3}^\mathrm{T}\,\mathbf{T}_{d}^\mathrm{T} \right]^\mathrm{T}\rightarrow \mathbf{e}_{\bm{\theta}_G}(\bullet)$.

\begin{figure}[]
	\centering
		\subfloat[] {\includegraphics[width=.8\columnwidth]{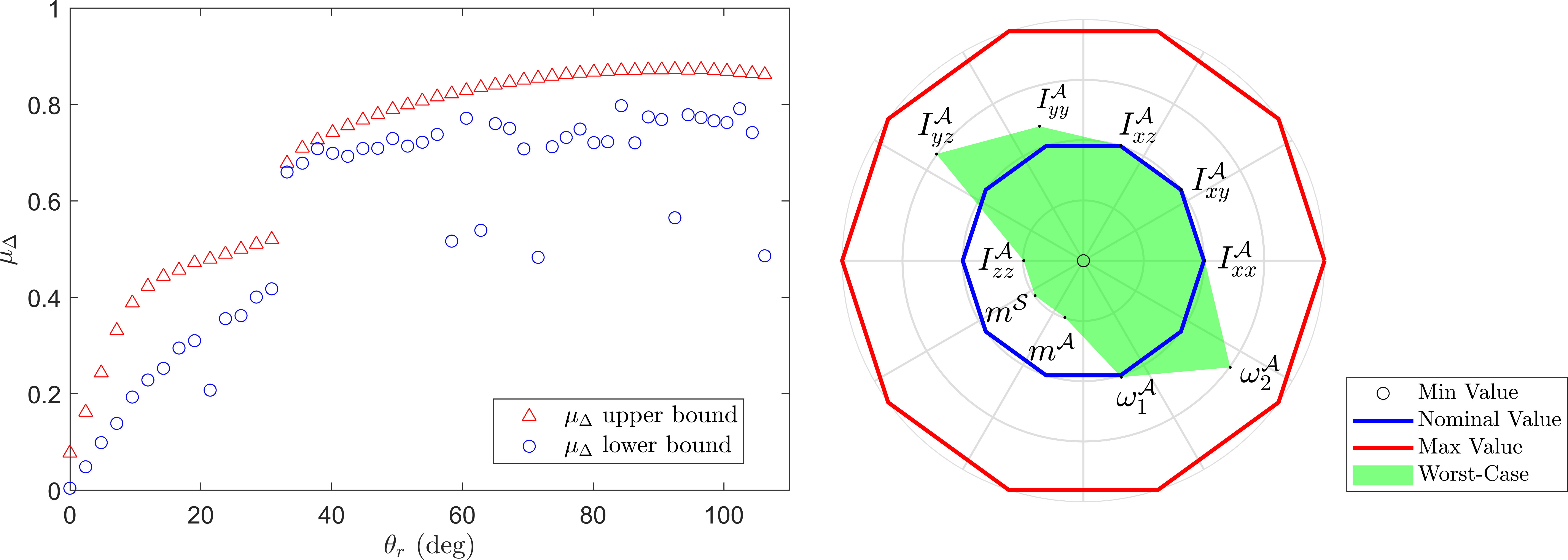}}\\ \vspace{2pt}
		\subfloat[] {\includegraphics[width=.8\columnwidth]{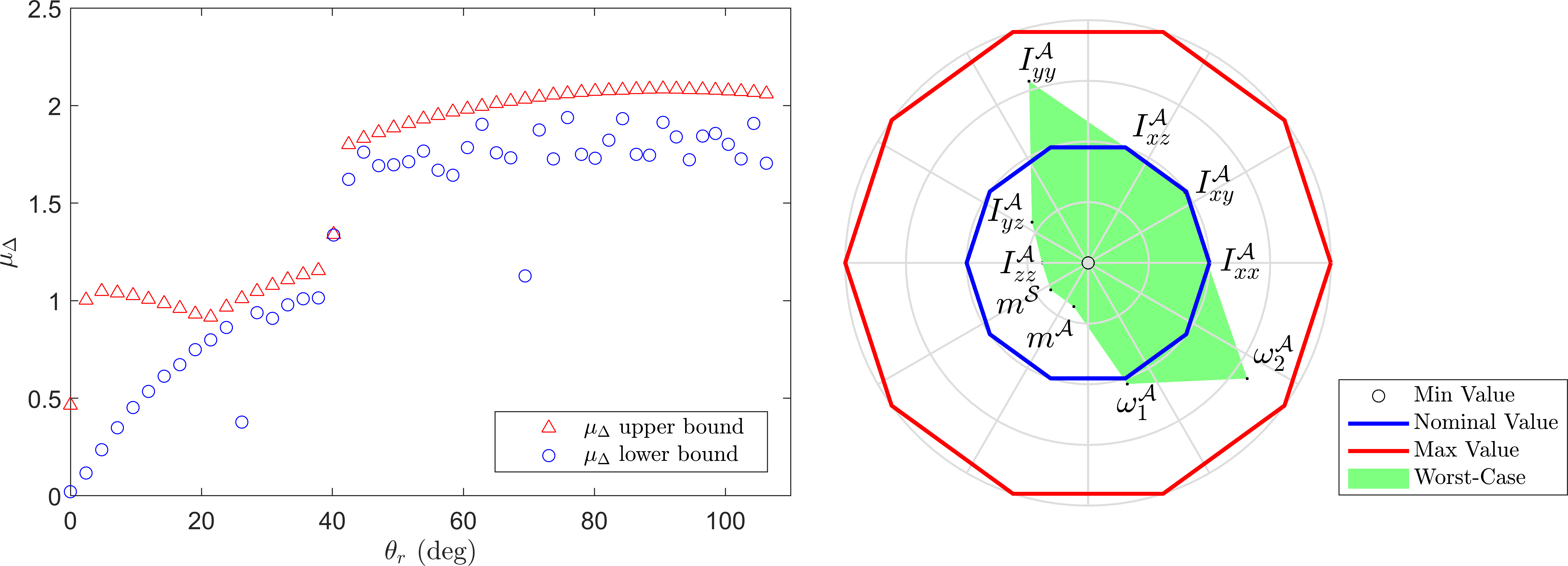}}
\caption{Structured singular value bounds and worst-case parameter combination for the transfer $\left[\mathbf{d}_{i_3}^\mathrm{T}\,\mathbf{T}_{d}^\mathrm{T} \right]^\mathrm{T}\rightarrow \mathbf{e}_{\bm{\theta}_G}(1)$: (a) slow rotor rate, (b) fast rotor rate }
\label{fig:polygon_slow_plus}
\end{figure}

This computation is efficiently provided by \texttt{wcgain} routine in \textsc{Matlab} \textit{Robust Control Toolbox} \cite{balas2016}. 
Let consider the set of parametric uncertainties in Table \ref{tab:wc_comb}. If we take as reference the transfer $\left[\mathbf{d}_{i_3}^\mathrm{T}\,\mathbf{T}_{d}^\mathrm{T} \right]^\mathrm{T}\rightarrow \mathbf{e}_{\bm{\theta}_G}(1)$ the upper and lower bounds of the structured singular values are shown in Fig. \ref{fig:polygon_slow_plus} for 49 values of $\tau \in \left[0,0.5\right]$. This interval is chosen since the model is symmetric in the intervals $\theta_r \in \left[0,90\right]^\circ$. For the same reason the results in the interval $\tau \in \left[-1,0\right]$ are not shown. The worst-case $\mu_\Delta$ bounds are presented in Table \ref{tab:wc_mu}.

\begin{table}
	\centering
	{\begin{tabular}{cccccc}
		\hline
		\textbf{Parameter} &\textbf{Nominal Value} &\textbf{Unit} & \textbf{Uncertainty} & \textbf{WC for} $\bm{\Omega_q = 0.06\,\mathrm{^\circ/s}}$ & \textbf{WC for} $\bm{\Omega_q = -0.205\,\mathrm{^\circ/s}}$ \\ \hline
		$m^\mathcal{S}$ & 1147 &$\mathrm{kg}$ & $20\%$ & 1031.15 & 1004.60\\
		$m^\mathcal{A}$ & 43.2 & $\mathrm{kg}$ & $20\%$ & 38.8367 & 37.8367 \\
		$I_{xx}^\mathcal{A}$ & 75.41 & $\mathrm{kg\, m^2}$ & $20\%$ & 75.5085 & 75.1802 \\
		$I_{yy}^\mathcal{A}$ & 21.86 & $\mathrm{kg\, m^2}$ & $20\%$ & 22.6100 & 24.3660 \\
		$I_{zz}^\mathcal{A}$ & 80.45 & $\mathrm{kg\, m^2}$ & $20\%$ &  72.3244 & 70.4622 \\
		$I_{xy}^\mathcal{A}$ & 0.43 & $\mathrm{kg\, m^2}$ & $20\%$ & 0.4303 & 0.4289\\
		$I_{yz}^\mathcal{A}$ & 22.04 & $\mathrm{kg\, m^2}$ & $20\%$ & 24.2661 & 20.1493 \\
		$I_{xz}^\mathcal{A}$ & -0.1 & $\mathrm{kg\, m^2}$ & $20\%$ & -0.1001 & -0.1001 \\
		$\omega_{1}^\mathcal{A}$ & 0.42 & $\mathrm{Hz}$ & $20\%$ & 0.4213 & 0.4241 \\
		$\omega_{2}^\mathcal{A}$ & 0.61 & $\mathrm{Hz}$ & $20\%$ & 0.6716 & 0.6857 \\ \hline
	\end{tabular}}
	\caption{Worst-case parameter combination}
	\label{tab:wc_comb}
\end{table}

\begin{table}
	\centering
	{\begin{tabular}{ccccc}
		\hline
		$\bm{\Omega_q\,(\mathrm{^\circ/s)}}$ & $\bm{\mathrm{inf}_\Delta \mu_\Delta}$ & $\bm{\mathrm{sup}_\Delta \mu_\Delta}$ & \textbf{Critical Frequency (Hz)} & \textbf{Critical rotor angle (deg)}\\ \hline
		0.06 & 0.7222 & 0.8723 & 0.9202 & 90 \\
		-0.205 & 1.9141 & 2.0855 & 0.9334 & 90 \\ \hline
	\end{tabular}}
\caption{Worst-case $\mu_\Delta$ scenarios}
\label{tab:wc_mu}
\end{table}

\begin{figure}[h!]
	\centering
	\includegraphics[width=.7\columnwidth]{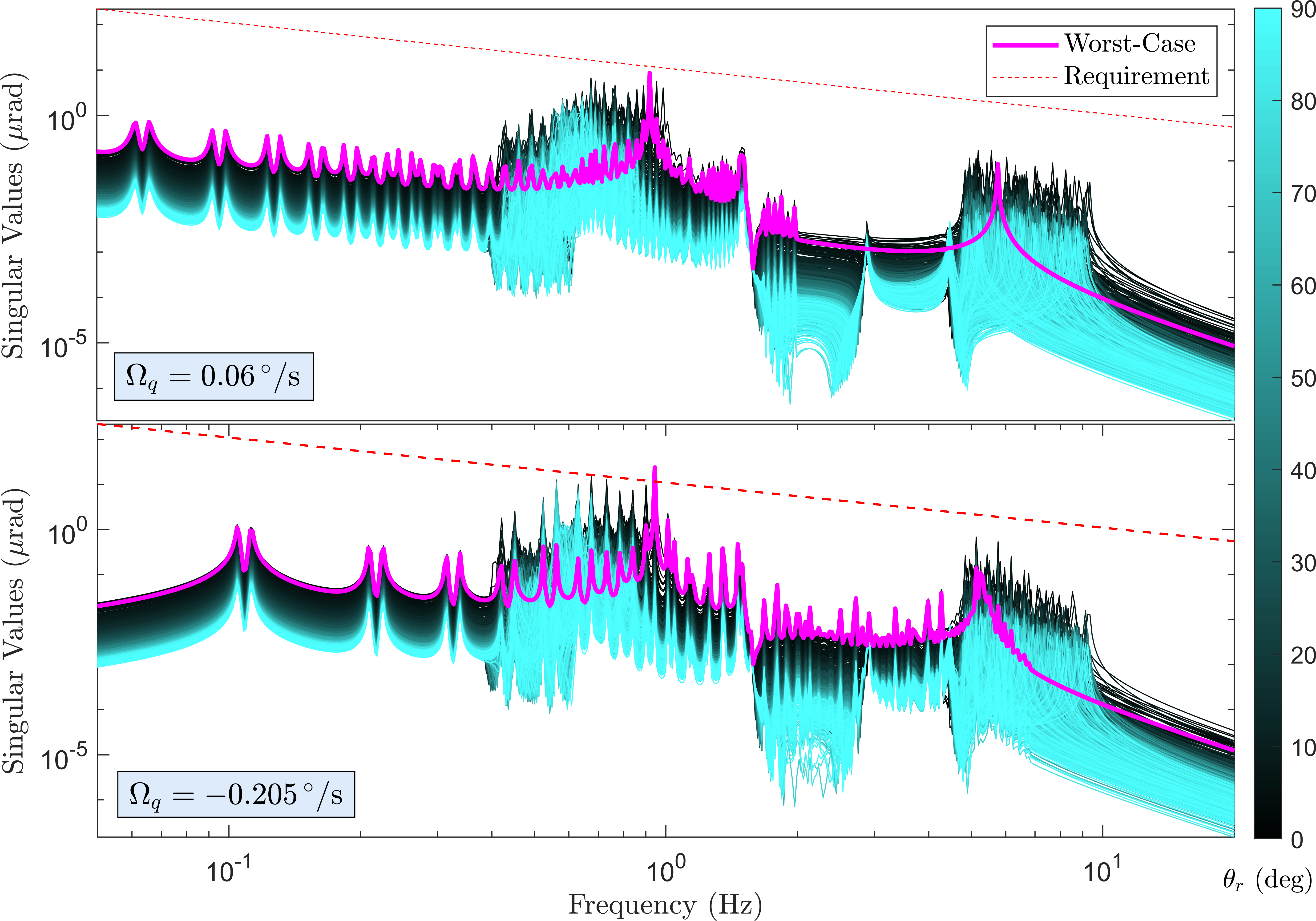} 
	\caption{Singular values of the transfer function $\mathbf{d}_{gb}\rightarrow\theta_G(1)$ as function of the rotor angular configuration compared with the worst-case scenario and the RPE pointing requirement for two cases: slow rotor rate (top), fast rotor rate (bottom)}
	\label{fig:sigma_slow_fast}
\end{figure}

From Fig. \ref{fig:polygon_slow_plus} we note that if for slow rotor rate the performance is met for any rotor angular configuration, this is not the case for fast rotor rate for most of the rotor configurations. In both cases the worst-case is reached for $\theta_r = 90\,\mathrm{^\circ}$ where the gearbox mode has maximum amplitude (see Fig. \ref{fig:transf_ustep_theta}). 
Figure \ref{fig:polygon_slow_plus} shows also the worst-case parameter combination normalized with the absolute maximum value of each uncertain parameter. From this analysis it is possible to detect the parameters which have the biggest impact on the degradation of the pointing performance. 
The worst-case parameter combination is also detailed in Table \ref{tab:wc_comb}.
Figure \ref{fig:sigma_slow_fast} finally shows the singular values of the uncertain transfer function $\mathbf{d}_{gb}\rightarrow\theta_G(1)$  as function of the rotor angular configuration together with the worst-case of the same transfer corresponding to the parameter combination in Table \ref{tab:wc_comb}. The RPE requirement ($W_{\mathrm{RPE}}^{-1}$) is also plotted in the same figure. Note as for $\Omega_q = -0.205\,\mathrm{^\circ/s}$ the worst-case curve breaks the requirement in correspondence of the critical frequency.

What we showed in this study is just an example of what it can be achieved using the TITOP multi-body approach. With the assembled model of any multi-body flexible structure in LFT form it is possible to detect worst-case scenarios without using time-consuming Monte-Carlo approaches classically used in industry. Compared to standard, point-wise Monte-Carlo analysis, the methodology proposed in this paper insures that the full uncertainty space is covered by the worst-case analysis. 

\section{On-line estimator of SADM disturbance}
\label{sec:observer}

In this section the model in Fig. \ref{fig:lpv_sadm_satellite} is used to design an LPV oberver scheduled with the rotation parameter $\tau$ able to estimate the source of micro-vibration directly on-board by combining the gyrometer and star tracker measurements used by the underlying ACS. 

\begin{figure}[h!]
	\centering
	\includegraphics[width=.8\columnwidth]{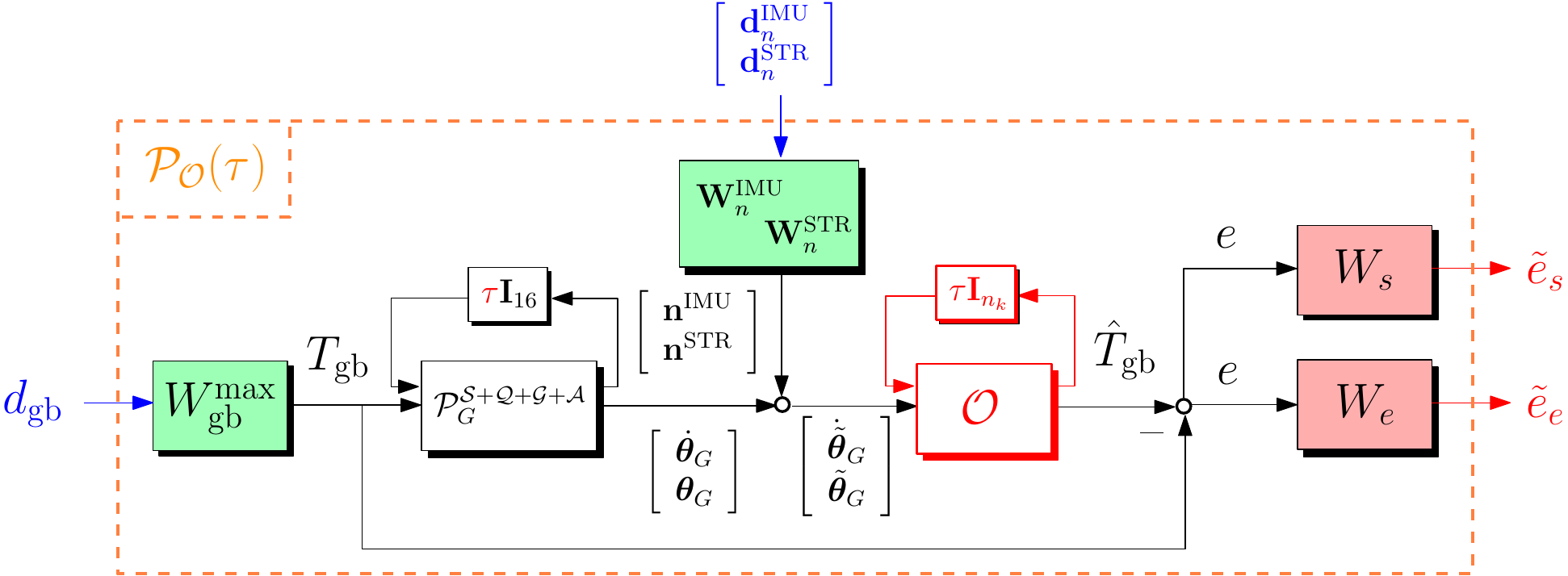} 
	\caption{LPV observer synthesis: connection diagram}
	\label{fig:h_inf_synthesis_nominal_plant}
\end{figure}

Let consider the diagram in Fig. \ref{fig:h_inf_synthesis_nominal_plant} used for the synthesis of the LPV observer $\mathcal{O}(\tau)$. Note that for this synthesis we consider the LPV nominal plant $\mathcal{P}_G^{\mathcal{S},\mathcal{Q},\mathcal{G},\mathcal{A}}(\tau)$ which describes the propagation from the gearbox disturbance torque to the three angular speeds $\dot{\bm{\theta}}_G$ and the three attitude angles ${\bm{\theta}}_G$. 

\paragraph{Input weights}
The input weights are of two types:
\begin{itemize}[noitemsep,topsep=0pt]
	\item Weight ${W}_\mathrm{gb}^\mathrm{max}$ which shapes the expected upper bounds of the amplitude spectrum of the harmonic input disturbance $T_{gb}$ (deterministic signal). The chosen value is $1.2/\pi$, taken by considering 20\% more than the maximum amplitude from \eqref{eq:gearbox_sawtooth}.
	\item Weights $\mathbf{W}_n^\mathrm{IMU}$ and $\mathbf{W}_n^\mathrm{STR}$ that shape respectively the amplitude spectral density of the IMU and the STR noises (stochastic signals).
	For this work $\mathbf{W}_n^\mathrm{IMU} = 0.03\,(\mathrm{\mu rad/s)/\sqrt{Hz}}$ and $\mathbf{W}_n^\mathrm{STR} = 1\,\mathrm{\mu rad/\sqrt{Hz}}$.
\end{itemize}

\paragraph{Output weights}
The output weights are of two types:
\begin{itemize}[noitemsep,topsep=0pt]
	\item Weight $W_e$ that shapes in frequency the upper bound of the amplitude spectrum of the maximum estimation error $e$ that is the difference between the estimated disturbance torque $\hat{T}_{gb}$ and the actual torque $T_{gb}$. 
	For this study it is chosen a 6th-order band-pass Butterworth filter which passes frequencies between $0.025$ and $0.74\,\mathrm{Hz}$ and demand a reduction of the error of $\approx 27\,\mathrm{dB}$.
	This filter is tuned in order to have a good estimation of the amplitudes and frequencies of the first 30 harmonics corresponding to the two gearbox imperfections $\tilde{\Omega}_d = 184$ and $\tilde{\Omega}_d = 197.3$.
	\item Weight $W_s$ that shapes the maximum variance of the estimation error $e$. The chosen value is $1/0.13$ in order to limit the noise amplification to $0.13$. 
\end{itemize}

The structure chosen for the LPV observer $\mathcal{O}(\tau)$ is:

\begin{equation}
	\mathcal{O}(\tau) = \mathcal{F}_u\left(\left[\begin{array}{c|c}
		\mathbf{A}_{\mathcal{O}_0} & \mathbf{B}_{\mathcal{O}_0} \\ \hline
		\mathbf{C}_{\mathcal{O}_0} & \mathbf{D}_{\mathcal{O}_0}
	\end{array}\right]+\tau\left[\begin{array}{c|c}
		\mathbf{A}_{\mathcal{O}_1} & \mathbf{B}_{\mathcal{O}_1} \\ \hline
		\mathbf{C}_{\mathcal{O}_1} & \mathbf{D}_{\mathcal{O}_1}
	\end{array}\right],\frac{\mathbf{I}_{n_c}}{s}\right)= \mathcal{F}_u\left({\mathbf{\mathcal{O}}},\tau\mathbf{I}_{n_k}\right)
\label{eq:obs}
\end{equation}

with $\mathcal{O} \subset \mathbb{R}\mathbb{H}_\infty^{(n_c+n_u)\times(n_c+n_y)}$. $\mathbb{R}\mathbb{H}_\infty^{(n_c+n_u)\times(n_c+n_y)}$ represents the set of finite gain transfer matrices with $(n_c+n_u)$ inputs and $(n_c+n_y)$ outputs. $n_c$ is the observer order, $n_u$ is the number of inputs, $n_y$ is the number of outputs and finally $n_k$ is the number of occurrences of the scheduling parameter $\tau$. The matrices
\begin{equation*}
	\mathbf{A}_{\mathcal{O}_0},\,\mathbf{A}_{\mathcal{O}_1},\,\mathbf{B}_{\mathcal{O}_0},\,\mathbf{B}_{\mathcal{O}_1},\,\mathbf{C}_{\mathcal{O}_0},\,\mathbf{C}_{\mathcal{O}_1},\,\mathbf{D}_{\mathcal{O}_0},\,\mathbf{D}_{\mathcal{O}_1}
\end{equation*}
 are real matrices of appropriate dimension.
 
 The observer is synthesized using the non-smooth optimization algorithms \cite{apk2014} available through the \textsc{Matlab} routine \texttt{systune}. In order to have satisfactory results it is necessary to find a good initial guess for the optimization algorithm. This is why the synthesis process consists of four steps detailed hereafter:

 	\textbf{Step 1:} Synthesize the full order unstructured observer $\mathcal{O}_\mathrm{opt}$ using standard $\mathcal{H}_\infty$ algorithms, i.e. solve the problem:
 	\begin{equation}
 		\underset{\mathcal{O}_\mathrm{opt}}{\mathrm{minimize}} \quad \gamma \quad s.t. \quad \left\vert\left\vert \mathcal{P}_{\mathcal{O}_{\mathbf{d}_{\mathrm{gb}}\rightarrow \tilde{e}_e}}\left( \tau_\mathrm{opt}\right)\right\vert\right\vert_\infty < \gamma,
 	\end{equation}
 where $\tau_\mathrm{opt}$ is chosen according to the worst-case configuration discussed in Section \ref{sec:wc_pointing_analysis}.
 
 \textbf{Step 2:} Reduce $\mathcal{O}_\mathrm{opt}$ by a balanced model reduction \cite{zhou1993frequency}. The resulting reduced observer $\mathcal{O}_\mathrm{red}$ is of order 4.
 
 \textbf{Step 3:} Synthesize the reduced-order observer $\tilde{\mathcal{O}}_\mathrm{red}$ using the non-smooth nominal controller design techniques \cite{apk2014} available in \textsc{Matlab} \texttt{systune} routine, i.e. solve the following mixed $\mathcal{H}_\infty/\mathcal{H}_2$ problem by initializing the optimization with $\mathcal{O}_\mathrm{red}$:
 \begin{equation}
 	\underset{\tilde{\mathcal{O}}_\mathrm{red}}{\mathrm{minimize}} \quad \gamma_1,\gamma_2 \quad s.t. \begin{cases}
 			 \vert\vert \mathcal{P}_{\mathcal{O}_{\mathbf{d}_{\mathrm{gb}}\rightarrow \tilde{e}_e}}( \tau_\mathrm{opt})\vert\vert_\infty < \gamma_1 \quad \mathrm{(hard\,constraint)}\\
 			 \vert\vert \mathcal{P}_{\mathcal{O}_{
 			 		\mathbf{d}_n\rightarrow \tilde{e}_s}}( \tau_\mathrm{opt})\vert\vert_2 < \gamma_2 \quad \mathrm{(soft\,constraint)}
 			\end{cases}
 		\label{eq:obs_mix}
 \end{equation}

\textbf{Step 4:} Let $\overline{\bm{\tau}}$ be a grid across the domain of the scheduling parameter $\bm{\tau}=\left[0,1\right]$.
With $\tilde{\mathcal{O}}_\mathrm{red}$ initialize the matrices $\mathbf{A}_{\mathcal{O}_0}$, $\mathbf{B}_{\mathcal{O}_0}$, $\mathbf{C}_{\mathcal{O}_0}$ and $\mathbf{D}_{\mathcal{O}_0}$ of the observer \eqref{eq:obs}. Consider $\tau$ as an uncertain parameter and optimize using the non-smooth parametric robust controller design technique proposed in \cite{Apkarian2015} by solving for each $\tau_j\in\overline{\bm{\tau}}$ the mixed $\mathcal{H}_\infty/\mathcal{H}_2$ problem:
\begin{equation}
	\underset{{\mathcal{O}}_j}{\mathrm{minimize}} \quad \gamma_1,\gamma_2 \quad s.t. \begin{cases}
		\vert\vert \mathcal{P}_{\mathcal{O}_{\mathbf{d}_{\mathrm{gb}}\rightarrow \tilde{e}_e}}( \tau_j)\vert\vert_\infty < \gamma_1 \quad \mathrm{(hard\,constraint)}\\
		\vert\vert \mathcal{P}_{\mathcal{O}_{
				\mathbf{d}_n\rightarrow \tilde{e}_s}}( \tau_j)\vert\vert_2 < \gamma_2 \quad \mathrm{(soft\,constraint)}
	\end{cases}
	\label{eq:obs_mix_fin}
\end{equation}

\begin{figure}[h!]
	\centering
	\includegraphics[width=\columnwidth]{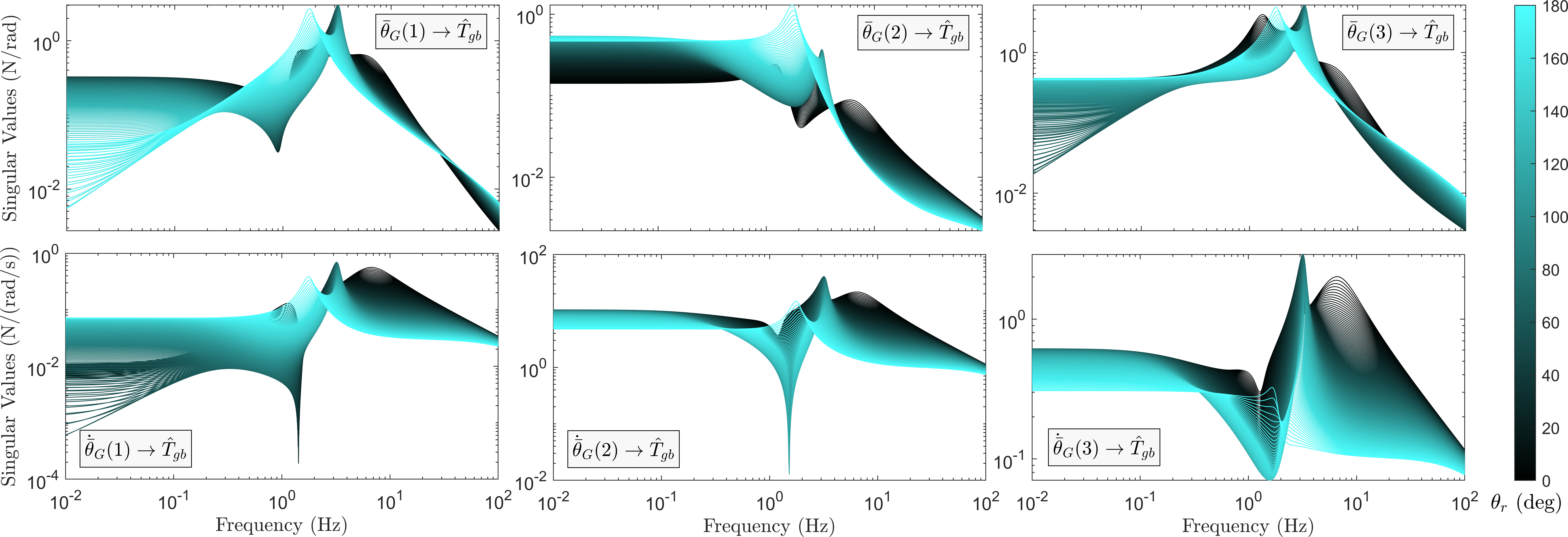} 
	\caption{LPV observer bode diagram}
	\label{fig:lpv_observer}
\end{figure}

With this procedure the resulting controller (see Fig. \ref{fig:lpv_observer}) has $n_c = 4$ states and $n_k = 5$ occurrences of the parameter $\tau$. It achieves a performance for the hard constraint of $\gamma_1 = 1.18$.

\begin{figure}[h!]
	\centering
	\includegraphics[width=.7\columnwidth]{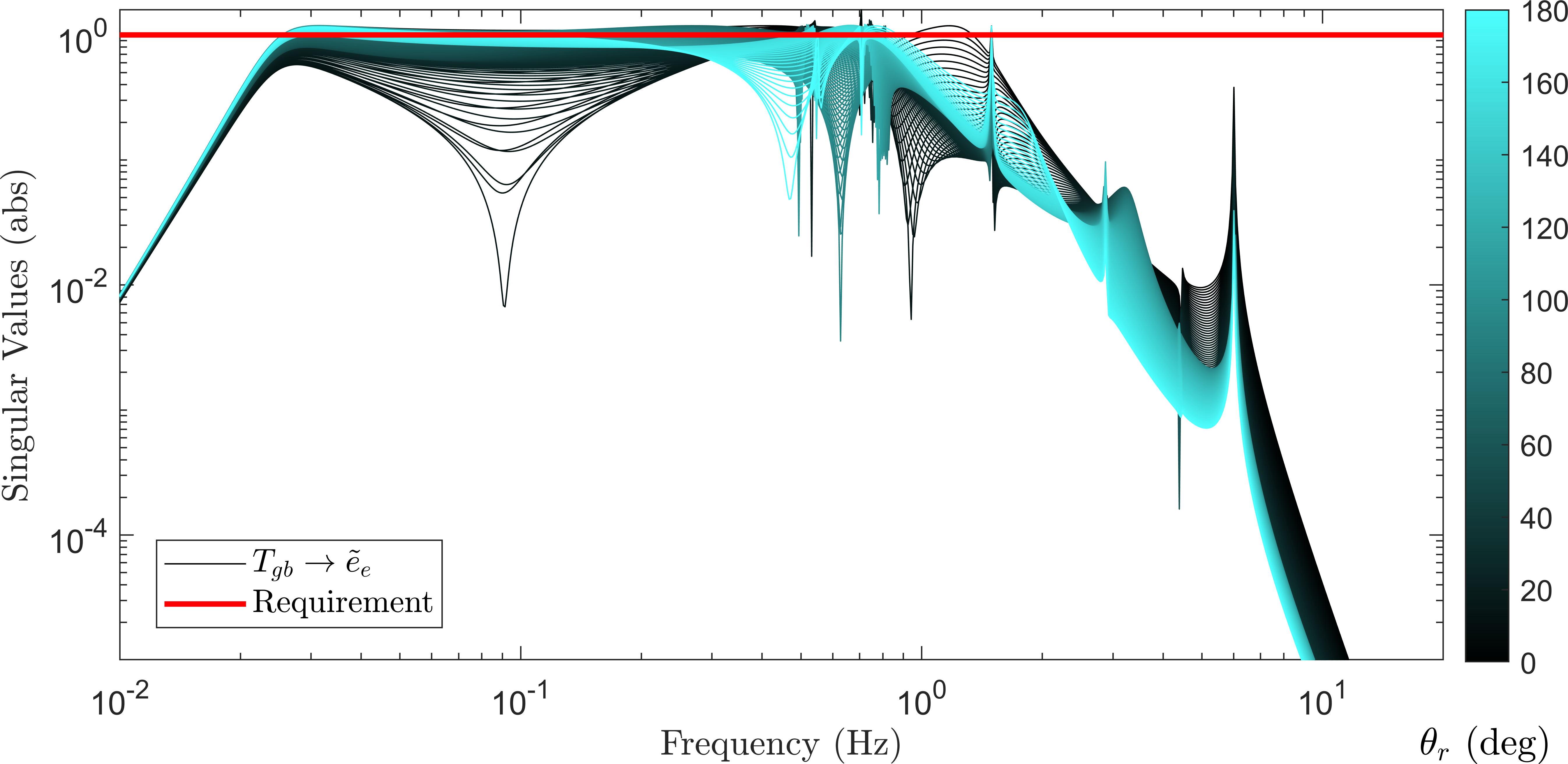} 
	\caption{Observer synthesis assessment. Transfer function $T_\mathrm{gb}\rightarrow \tilde{e}_e$}
	\label{fig:closed_loop_assessment}
\end{figure}

The singular values plot of the transfer function $T_\mathrm{gb}\rightarrow \tilde{e}_e$ is shown in Fig. \ref{fig:closed_loop_assessment}.

The soft constraint is imposed in order to limit the variance of the transfer $\mathbf{d}_n\rightarrow \tilde{e}_s$ and provide a good estimation of the disturbance signal $T_{gb}$ also in time domain.

\begin{figure}[h!]
	\centering
	\includegraphics[width=.7\columnwidth]{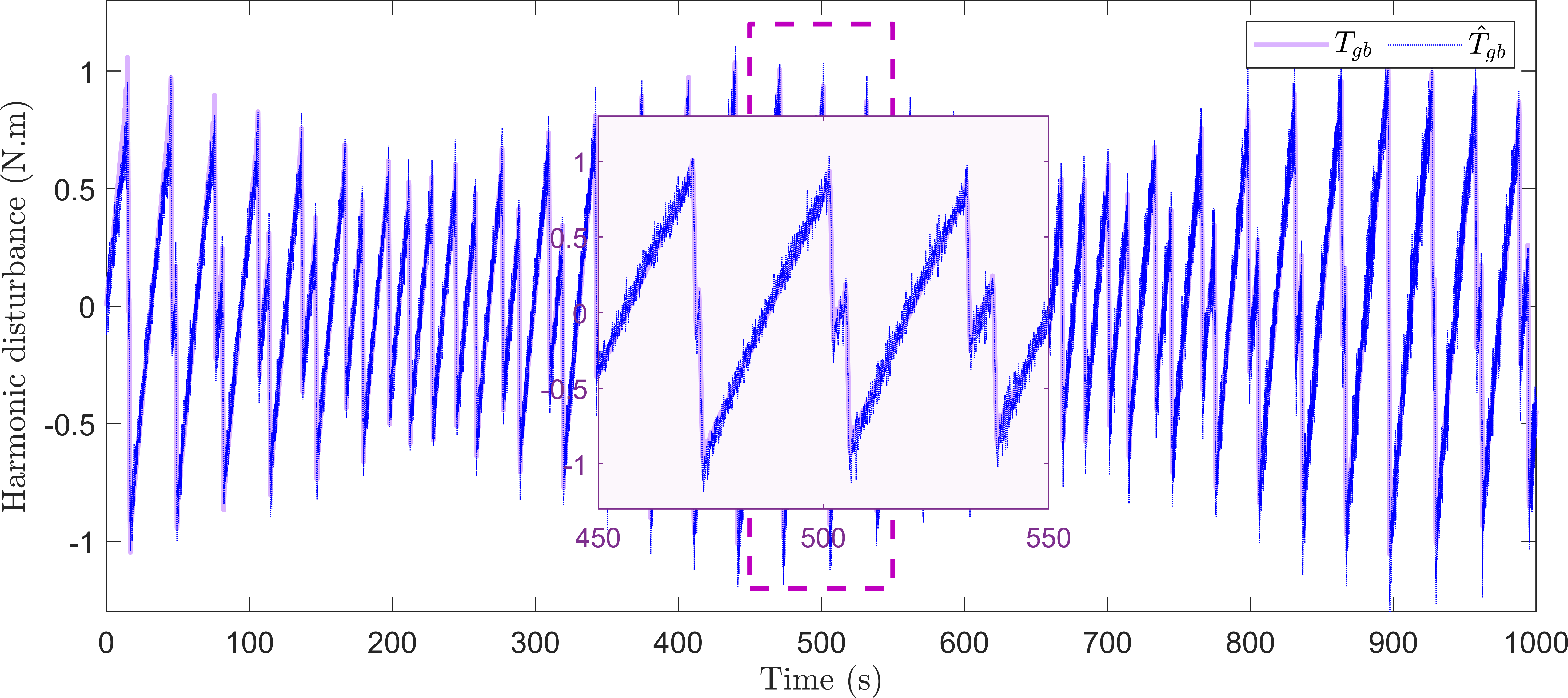} 
	\caption{Temporal simulation of the estimated gearbox disturbances}
	\label{fig:temporal_harmonic_est}
\end{figure}

A simulation of the system with 30 harmonics of the two gearbox imperfections $\tilde{\Omega}_d = 184$ and $\tilde{\Omega}_d = 197.3$ is performed for 1000 s for $\Omega_q = 0.06^\circ/s$ and $\tau(0) = 0$. The results of the estimation is shown in Fig. \ref{fig:temporal_harmonic_est}. The synthesized observer is able to capture in time domain the disturbance signal $T_{gb}$ with the evolution of the rotor configuration across the time.

\begin{figure}[h!]
	\centering
	\includegraphics[width=.7\columnwidth]{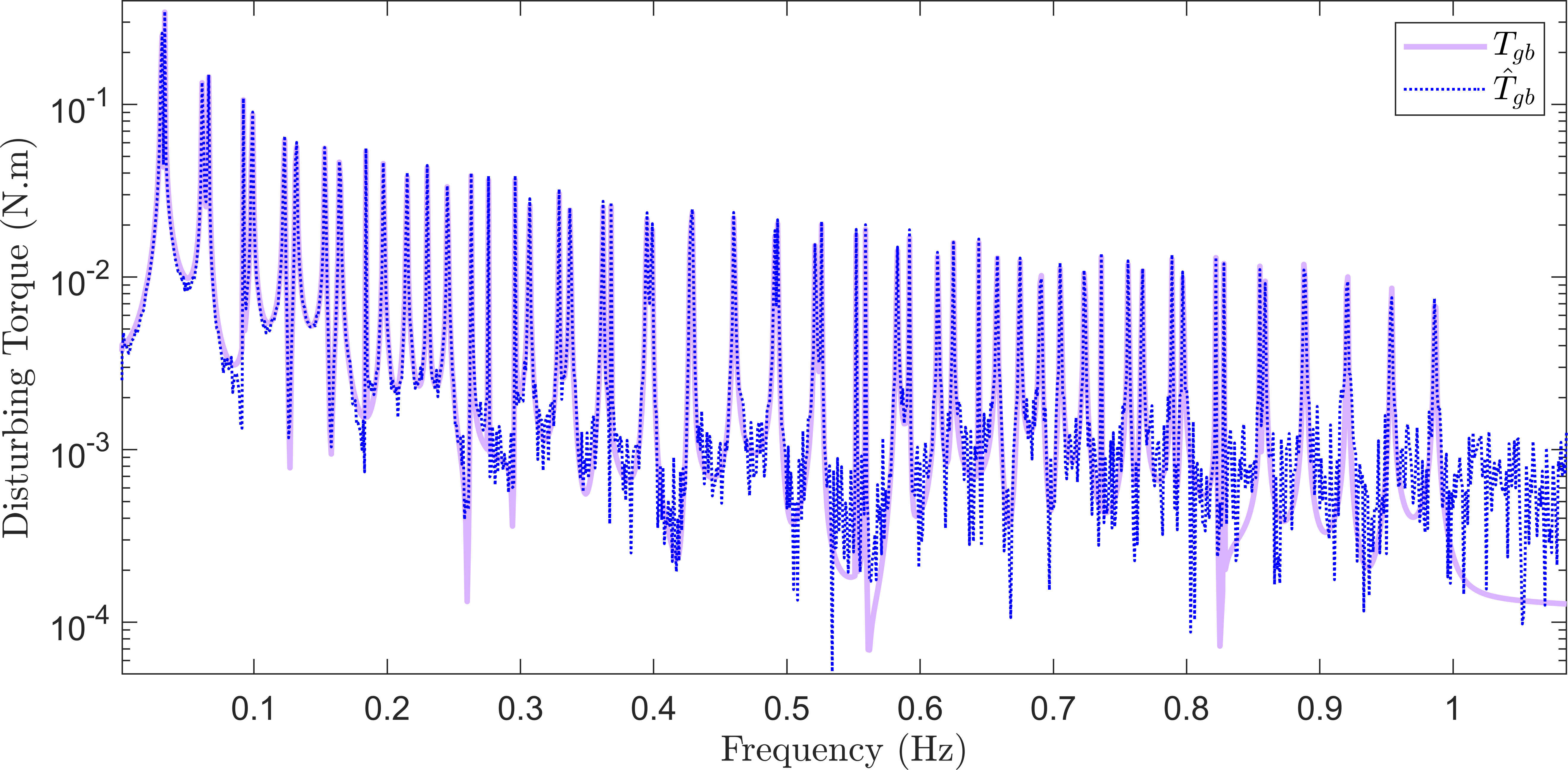} 
	\caption{FFT of the estimated gearbox disturbance}
	\label{fig:FFT_estimation}
\end{figure}

Figure \ref{fig:FFT_estimation} finally shows the FFT of the disturbance signal $T_{gb}$ and its estimate $\hat{T}_{gb}$. All the peaks of the micro-vibration source are perfectly captured in amplitude and frequency.

\section{Conclusions}
\label{sec:conclusion}
This paper presented a novel and complete methodology to model, assess and estimate on-board the micro-vibrations induced by a solar array drive mechanism. After building the entire dynamic model of a flexible spacecraft using the Two-Input Two-Output Ports (TITOP) approach including an accurate electromagnetic and mechanical model of the SADM, a validation in time and frequency domain is provided thanks to the telemetries of a European observation platform.

The model including a set of uncertainties is then directly used to robustly assess the pointing performance. This analysis allowed making predictions on worst-case scenarios in preliminary design phase by obtaining the corresponding worst-case parameter combination and critical solar array configurations.

Finally an LPV observer is designed and scheduled according to the solar array angular configuration in order to estimate directly on-board the disturbances induced by the SADM gearbox imperfections thanks to the measurements available in a classical attitude control system. 

The final aim of this work is to show how the presented TITOP model-based approach can handle the difficult task of making predictions of pointing performance in preliminary design phase. This approach conserves the physical and uncertain nature of the problem and directly monitor the impact of the design mechanical parameters without relaying on classical time-consuming simulation-based approaches (like Monte Carlo campaigns). 

\section*{Acknowledgement}
This research was supported by ESA (grant NO. REP/3-16604/20/NL/CRS/vr).

    \appendix
    
    \section{Spacecraft mechanical data for the study case}
    \label{tab:data}
    \begin{table}[h!]
    	\label{tab:spacecraft_data_sadm}
    	\centering
    	\begin{footnotesize}
    		\resizebox{\textwidth}{!}{\begin{tabular}{cllr}
    				\hline
    				\textbf{System} & \textbf{Parameter} & \textbf{Description} & \textbf{Nominal Value } \\ \hline
    				& $m^{\mathcal{S}}$ & Mass & $1147\,\mathrm{kg}$\\
    				& $\left[\begin{array}{ccc}
    					I_{xx}^{\mathcal{S}} & I_{xy}^{\mathcal{S}} & I_{xz}^{\mathcal{S}} \\
    					& I_{yy}^{\mathcal{S}} & I_{yz}^{\mathcal{S}} \\
    					sym & & I_{zz}^{\mathcal{S}}
    				\end{array}\right]$ & Inertia in $\mathcal{R}_s$ frame & $\left[\begin{array}{ccc}
    					381 & -9.4 & 3.1 \\
    					& 1015 & -38 \\
    					sym & & 935
    				\end{array}\right]\mathrm{kg\, m^2}$ \\
    				& $\mathbf{r}_G$ & Spacecraft CG & $\left[1.373\,\,0.014\,\,0.031\right]\,\mathrm{m}$\\
    				\multirow{-6}{*}{\shortstack{Central Body \\$\mathcal{S}$}} & $\mathbf{r}_P$ & $\mathcal{A}$ attachment point &  $\left[1.250\,\,-0.625\,\,0\right]\,\mathrm{m}$\\ \hline
    				& $m^{\mathcal{A}}$ & Mass & $43.2\,\mathrm{kg}$\\
    				&  $\left[\begin{array}{ccc}
    					I_{xx}^{\mathcal{A}} & I_{xy}^{\mathcal{A}} & I_{xz}^{\mathcal{A}} \\
    					& I_{yy}^{\mathcal{A}} & I_{yz}^{\mathcal{A}} \\
    					sym & & I_{zz}^{\mathcal{A}}
    				\end{array}\right]$ & Inertia in $\mathcal{R}_a$ frame &  $\left[\begin{array}{ccc}
    					75.41& 0.43 & -0.1 \\
    					& 21.86 & 22.04 \\
    					sym & & 80.45
    				\end{array}\right]\mathrm{kg\, m^2}$\\
    				& $\mathbf{r}_{o}^{\mathcal{A}}$ & $\mathcal{A}$ CG in $\mathcal{R}_a$ frame &  $\left[-0.002\,\,-1.745\,\,0.030\right]\,\mathrm{m}$\\
    				& $\left[\omega_1^{\mathcal{A}}\, \omega_2^{\mathcal{A}}\, \omega_3^{\mathcal{A}}\, \omega_4^{\mathcal{A}}\, \omega_5^{\mathcal{A}}\right]$ & Flexible modes' frequencies & $\left[0.42\,\, 0.61\,\,1.58\,\, 2.83\,\, 4.30 \right]\,\mathrm{Hz}$ \\
    				& $\zeta_1^{\mathcal{A}}, \zeta_2^{\mathcal{A}}, \zeta_3^{\mathcal{A}}, \zeta_4^{\mathcal{A}}, \zeta_5^{\mathcal{A}}$ & Flexible modes' damping & $0.03$\\
    				\multirow{-10}{*}{\shortstack{Solar \\ Array $\mathcal{A}$}} & $\mathbf{L}_P^{\mathcal{A}}$ & Modal participation factors & $\left[\begin{array}{ccccc}
    					0.0019 & 5.0012 & 0.3244 & 0.0004 & 0.0053 \\
    					-1.5151 & 0.0033 & 0.0019 & -0.9132 & 5.8780 \\
    					-4.7387 & -0.0042 & 0.0288 & 2.7900 & -1.0144 \\
    					13.9436 & 0.0093 & -0.0235 & -1.8763 & 3.0180 \\
    					0.0229 & 1.8457 & 3.2879 & -0.0246 & -0.0124 \\
    					-0.0144 & 14.5319 & -0.2128 & 0.0031 & -0.0001
    				\end{array} \right]^{\mathrm{T}}$\\ \hline
    				& $K_m$ & Motor torque const & $4.44\,\mathrm{N\, m/(A\, rad)}$\\
    				& $z$ & Rotor teeth & 90 \\
    				& $p$ & Motor poles & 4\\
    				& $n_{\mu}$ & $\mu$-step subdivisions & 8\\
    				& $I$ & Phase current & $0.17\,\mathrm{A}$ \\
    				& $J_r$ & Rotor inertia & $1.1\cdot 10^{-4}\,\mathrm{kg\, m^2}$ \\
    				& $C_r$ & Viscous damping & $13\cdot 10^{-4}\mathrm{N\, m\, s/rad}$ \\
    				& $\Omega_{q}$ & Slow rotor angular speed & $0.06\,\mathrm{^\circ/s}$\\
    				\multirow{-9}{*}{\shortstack{Solar \\ Array \\ Drive \\ Mechanism\\$\mathcal{Q}$}} &  & Fast rotor angular speed & $-0.205\,\mathrm{^\circ/s}$\\ \hline
    				& $J_o$ & Output shaft inertia & $0.01\,\mathrm{kg\, m^2}$ \\
    				& $K_g$ & Stiffness & $9600\,\mathrm{N\, m/rad}$ \\
    				& $C_g$ & Viscous damping & $0.1\,\mathrm{N\, m\, s/rad}$ \\
    				\multirow{-3}{*}{Gearbox $\mathcal{G}$} & $N_g$ & Reduction ratio & 184\\ \hline 
    		\end{tabular}}
    	\end{footnotesize}
    	\caption{Spacecraft data}
    \end{table}
    
    \section{TITOP State-Space Models of a SADM}
    
    \begin{small}
    \begin{equation}
    	\left[
    	\begin{array}{c}
    		\dot{\theta}_r \\
    		\ddot{\theta}_r \\ \hline
    		\ddot{\mathbf{q}}_{\mathrm{out}}^{\mathcal{Q}} \\
    		\mathbf{W}_{\mathcal{Q}/\mathcal{S},P} \\ \hline
    		\theta_r
    	\end{array}
    	\right]=
    	\underbrace{
    		\left[
    		\begin{array}{c|c|c}
    			\mathbf{A} & \mathbf{B}_1 & \mathbf{B}_2 \\ \hline
    			\mathbf{C}_1 & \mathbf{D}_{11} & \mathbf{D}_{12} \\ \hline
    			\mathbf{C}_2 & \mathbf{D}_{21} & \mathbf{D}_{22}
    		\end{array}
    		\right]}_{\mathcal{M}_P^{\mathcal{Q}}}
    	\left[
    	\begin{array}{c}
    		\theta_r \\
    		\dot{\theta}_r \\ \hline
    		\mathbf{W}_{\mathcal{A}/\mathcal{Q},P} \\
    		\ddot{\mathbf{q}}_{\mathrm{in}}^{\mathcal{Q}} \\ \hline
    		i
    	\end{array}
    	\right],
    	\label{eq:sadm_titop}
    \end{equation}
    where
    
    \begin{equation*}
\mathbf{A} = \left[
\begin{array}{cc}
	0 & 1 \\
	-\frac{K_0}{J_r} & -\frac{C_r}{J_r}
\end{array}
\right], \quad
\mathbf{B_1} = \left[
\begin{array}{cc}
	\mathbf{0}_{1\times 6} & \mathbf{0}_{1\times 6} \\
	\left[\begin{array}{cc}
		\mathbf{0}_{1\times 5} & -\frac{1}{J_r} 
	\end{array}\right] & \left[ 
	\begin{array}{cc}
		\mathbf{0}_{1\times 5} & \frac{1}{J_r} 
	\end{array}
	\right] 
\end{array}
\right], \quad
\mathbf{B}_2 = \left[
\begin{array}{c}
	0 \\
	K_m I\gamma
\end{array}
\right]
    \end{equation*}
    
    \begin{equation*}
	\mathbf{C_1} = \left[
\begin{array}{cc}
	\left[
	\begin{array}{c}
		\mathbf{0}_{5\times 1} \\
		-\frac{K_0}{J_r}
	\end{array}
	\right] & \left[
	\begin{array}{c}
		\mathbf{0}_{5\times 1} \\
		-\frac{C_r}{J_r}
	\end{array}
	\right] \\
	\left[\begin{array}{c}
		\mathbf{0}_{5\times 1} \\
		K_0
	\end{array}\right] & \left[\begin{array}{c}
		\mathbf{0}_{5\times 1} \\
		C_r
	\end{array}\right]
\end{array}
\right], \quad
\mathbf{D}_{11} = \left[
\begin{array}{cc}
	\left[
	\begin{array}{c}
		\mathbf{0}_{5\times 6} \\
		\begin{array}{cc}
			\mathbf{0}_{1\times 5} & \frac{1}{J_r}
		\end{array}
	\end{array}
	\right] & \left[
	\begin{array}{cc}
		\mathbf{I}_5 & \mathbf{0}_{5\times 1} \\
		\mathbf{0}_{1\times 5} & 1-\frac{1}{J_r}
	\end{array}
	\right] \\
	\left[
	\begin{array}{c}
		\begin{array}{cc}
			\mathbf{I}_{5} & \mathbf{0}_{5\times 1}
		\end{array} \\
		\mathbf{0}_{1\times 6}
	\end{array}
	\right] & \mathbf{0}_{6\times 6}
\end{array}
\right]
    \end{equation*}

\begin{equation*}
	\mathbf{C}_2 = \left[\begin{array}{cc}
		1 & 0
	\end{array}\right], \quad
	\mathbf{D}_{12} = \left[
	\begin{array}{c}
		-K_m I\gamma \\
		\left[
		\begin{array}{c}
			\mathbf{0}_{5\times 1} \\
			K_m I \gamma
		\end{array}
		\right]
	\end{array}
	\right], \quad
	\mathbf{D}_{21} = \mathbf{0}_{1\times 12}, \quad
	\mathbf{D}_{22} = 0
\end{equation*}
\end{small}

        \section{TITOP State-Space Models of a Reduction Gearbox}
                \begin{small}
        \begin{equation}
        	\left[
        	\begin{array}{c}
        		\dot{\delta\theta}_o \\
        		\ddot{\delta\theta}_o \\ \hline
        		\ddot{\mathbf{q}}_{\mathrm{out}}^{\mathcal{G}} \\
        		\mathbf{W}_{\mathcal{G}/\mathcal{Q},P} \\ \hline
        		\delta\theta_o
        	\end{array}
        	\right]=
        	\underbrace{
        		\left[
        		\begin{array}{c|c|c}
        			\mathbf{A} & \mathbf{B}_1 & \mathbf{B}_2 \\ \hline
        			\mathbf{C}_1 & \mathbf{D}_{11} & \mathbf{D}_{12} \\ \hline
        			\mathbf{C}_2 & \mathbf{D}_{21} & \mathbf{D}_{22}
        		\end{array}
        		\right]}_{\mathcal{M}_P^{\mathcal{G}}}
        	\left[
        	\begin{array}{c}
        		\delta\theta_o \\
        		\dot{\delta\theta}_o \\ \hline
        		\mathbf{W}_{\mathcal{A}/\mathcal{G},P} \\
        		\ddot{\mathbf{q}}_{\mathrm{in}}^{\mathcal{G}} \\ \hline
        		T_{gb}
        	\end{array}
        	\right],
        	\label{eq:gb_titop}
        \end{equation}
        where
        
        \begin{equation*}
	\mathbf{A} = \left[
\begin{array}{cc}
	0 & 1 \\
	-\frac{K_g}{J_o} & -\frac{C_g}{J_o}
\end{array}
\right], \quad
\mathbf{B_1} = \left[
\begin{array}{cc}
	\mathbf{0}_{1\times 6} & \mathbf{0}_{1\times 6} \\
	\left[\begin{array}{cc}
		\mathbf{0}_{1\times 5} & -\frac{1}{J_o} 
	\end{array}\right] & 
	\mathbf{0}_{1\times 6} 
\end{array}
\right], \quad
\mathbf{B}_2 = \left[
\begin{array}{c}
	0 \\
	\frac{1}{J_0}\gamma
\end{array}
\right]
        \end{equation*}
    \begin{equation*}
\mathbf{C_1} = \left[
\begin{array}{cc}
	\left[
	\begin{array}{c}
		\mathbf{0}_{5\times 1} \\
		-\frac{K_g}{J_o}
	\end{array}
	\right] & \left[
	\begin{array}{c}
		\mathbf{0}_{5\times 1} \\
		-\frac{C_g}{J_o}
	\end{array}
	\right] \\
	\left[\begin{array}{c}
		\mathbf{0}_{5\times 1} \\
		\frac{K_g}{N_g}
	\end{array}\right] & \left[\begin{array}{c}
		\mathbf{0}_{5\times 1} \\
		\frac{C_g}{N_g}
	\end{array}\right]
\end{array}
\right], \quad
\mathbf{D}_{11} = \left[
\begin{array}{cc}
	\left[
	\begin{array}{c}
		\mathbf{0}_{5\times 6} \\
		\begin{array}{cc}
			\mathbf{0}_{1\times 5} & \frac{1}{J_o}
		\end{array}
	\end{array}
	\right] &  \mathbf{0}_{6\times6} \\
	\left[
	\begin{array}{c}
		\begin{array}{cc}
			\mathbf{I}_{5} & \mathbf{0}_{5\times 1}
		\end{array} \\
		\mathbf{0}_{1\times 6}
	\end{array}
	\right] & \left[\begin{array}{c}
		\mathbf{0}_{5\times 6}  \\
		-J_i 
	\end{array}\right]
\end{array}
\right]
    \end{equation*}

\begin{equation*}
\mathbf{C}_2 = \left[\begin{array}{cc}
	1 & 0
\end{array}\right], \quad
\mathbf{D}_{12} = \left[
\begin{array}{c}
	\left[\begin{array}{c}
		\mathbf{0}_{5\times 1} \\
		\frac{1}{J_o}
	\end{array}\right] \\
	\left[
	\begin{array}{c}
		\mathbf{0}_{5\times 1} \\
		-\frac{1}{N_g}
	\end{array}
	\right]
\end{array}
\right], \quad
\mathbf{D}_{21} = \mathbf{0}_{1\times 12}, \quad
\mathbf{D}_{22} = 0.
\end{equation*}
\end{small}

\bibliographystyle{elsarticle-num-no-url} 
\bibliography{ms}%

\end{document}